\documentclass[fleqn,usenatbib]{mnras}

\usepackage[T1]{fontenc}
\usepackage{ae,aecompl}

\usepackage{graphicx}	
\usepackage{amsmath}	
\usepackage{amssymb}	

\usepackage[dvipsnames]{xcolor}
\usepackage{booktabs, blkarray}
\usepackage{mn2e-breakabs}
\usepackage{newtxtext,newtxmath}

\newcommand\Mpc{\, \mathrm{Mpc}}
\newcommand\Mpch{\, \mathrm{h}^{-1}\mathrm{Mpc}}
\newcommand\Mpchc{\, (\mathrm{Mpc/h})^{3}}
\newcommand\hMpc{\, \mathrm{h}\cdot \mathrm{Mpc}^{-1}}

\newcommand\Gpch{\, \mathrm{h}^{-1}\mathrm{Gpc}}

\newcommand{\bare}[1]{\!\!\left.\right|_\mathrm{#1}}

\newcommand\fse{f(z_{\rm eff})\sigma_8(z_{\rm eff})}
\newcommand\dmr{D_{\rm M}(z_{\rm eff})/r_{\rm drag}}
\newcommand\dhr{D_{\rm H}(z_{\rm eff})/r_{\rm drag}}

\definecolor{color27}{rgb}{0.2, 0.6, 0.8}

\usepackage[dvipsnames]{xcolor}
\usepackage[normalem]{ulem}

\title[eBOSS DR16 quasar sample Fourier Space analysis]{The Completed SDSS-IV extended Baryon Oscillation Spectroscopic Survey:
BAO and RSD measurements from the anisotropic power spectrum of the Quasar 
sample between redshift 0.8 and 2.2}

\author[R. Neveux et al.]{\parbox{\textwidth}{
Richard~Neveux\thanks{E-mail: richard.neveux@cea.fr}$^{1}$,
Etienne Burtin$^{1}$,
Arnaud de Mattia$^{1}$,
Alex Smith$^{1}$,
Ashley J. Ross$^{2}$,
Jiamin Hou$^{3}$,
Julian Bautista$^{4}$,
Jonathan Brinkmann$^{5}$,
Chia-Hsun Chuang$^{6}$,
Kyle~S.~Dawson$^{7}$,
H\'ector Gil-Mar\'in$^{8,9}$,
Brad W.\ Lyke$^{10}$,
Axel de la Macorra$^{11}$,
H\'elion~du~Mas~des~Bourboux$^{7}$,
Faizan G. Mohammad$^{12,13}$,
Eva-Maria M\"uller$^{14}$,
Adam D. Myers$^{10}$,
Jeffrey A. Newman$^{15}$,
Will J. Percival$^{12,13,16}$,
Graziano Rossi$^{17}$,
Donald Schneider$^{18}$,
M. Vivek$^{19,20}$,
Pauline Zarrouk$^{1,21}$,
Cheng Zhao$^{22}$,
Gong-Bo Zhao$^{23}$
} \vspace*{4pt} \\
\scriptsize $^{1}$ IRFU,CEA, Universit\'e Paris-Saclay, F-91191 Gif-sur-Yvette, France\vspace*{-2pt} \\ 
\scriptsize $^{2}$ Center for Cosmology and Astro-Particle Physics, Ohio State University, Columbus, Ohio, USA\vspace*{-2pt} \\ 
\scriptsize $^{3}$ Max-Planck-Institut f\"ur Extraterrestrische Physik, Postfach 1312, Giessenbachstr., 85748 Garching bei M\"unchen, Germany\vspace*{-2pt} \\ 
\scriptsize $^{4}$ Institute of Cosmology \& Gravitation, Dennis Sciama Building, University of Portsmouth, Portsmouth, PO1 3FX, UK\vspace*{-2pt} \\
\scriptsize $^{5}$ Apache Point Observatory, P.O. Box 59, Sunspot, NM 88349\vspace*{-2pt} \\
\scriptsize $^{6}$ Kavli Institute for Particle Astrophysics and Cosmology, Stanford University, 452 Lomita Mall, Stanford, CA 94305, USA\vspace*{-2pt} \\ 
\scriptsize $^{7}$ Department Physics and Astronomy, University of Utah, 115 S 1400 E, Salt Lake City, UT 84112, USA\vspace*{-2pt} \\ 
\scriptsize $^{8}$ Institut de Ci\`encies del Cosmos,  Universitat  de  Barcelona,  ICCUB,  Mart'i  i  Franqu\`es  1,  E08028  Barcelona,  Spain\vspace*{-2pt} \\
\scriptsize $^{9}$ Institut  d'Estudis  Espacials  de  Catalunya  (IEEC),  E08034  Barcelona,  Spain\vspace*{-2pt} \\ 
\scriptsize $^{10}$ University of Wyoming, 1000 E. University Ave., Laramie, WY 82071, USA\vspace*{-2pt} \\ 
\scriptsize $^{11}$ Instituto de F\'isica, Universidad Nacional Aut\'onoma de M\'exico, Apdo. Postal 20-364, M\'exico\vspace*{-2pt} \\ 
\scriptsize $^{12}$ Waterloo Centre for Astrophysics, University of Waterloo, Waterloo, ON N2L 3G1, Canada\vspace*{-2pt} \\ 
\scriptsize $^{13}$ Department of Physics and Astronomy, University of Waterloo, Waterloo, ON N2L 3G1, Canada\vspace*{-2pt} \\ 
\scriptsize $^{14}$ Sub-department of Astrophysics, Department of Physics, University of Oxford, Denys Wilkinson Building, Keble Road, Oxford OX1 3RH\vspace*{-2pt}  \\
\scriptsize $^{15}$ PITT PACC, Department of Physics and Astronomy, University of Pittsburgh, Pittsburgh, PA 15260, USA\vspace*{-2pt} \\ 
\scriptsize $^{16}$ Perimeter Institute for Theoretical Physics, 31 Caroline St. North, Waterloo, ON N2L 2Y5, Canada\vspace*{-2pt} \\ 
\scriptsize $^{17}$ Department of Physics and Astronomy, Sejong University, Seoul 143-747, Korea\vspace*{-2pt} \\ 
\scriptsize $^{18}$ Institute for Gravitation and the Cosmos, Pennsylvania State University, University Park, PA 16802, USA\vspace*{-2pt} \\ 
\scriptsize $^{19}$ Department of Astronomy \& Astrophysics, Pennsylvania State University, University Park, PA 16802, USA\vspace*{-2pt} \\ 
\scriptsize $^{20}$ Indian Institute of Astrophysics, Koramangala, Bangalore 560034, India\vspace*{-2pt} \\ 
\scriptsize $^{21}$ Institute for Computational Cosmology, Dept. of Physics, Univ. of Durham, South Road, Durham DH1 3LE, UK\vspace*{-2pt} \\
\scriptsize $^{22}$ Institute of Physics, Laboratory of Astrophysics, \'Ecole Polytechnique F\'ed\'erale de Lausanne (EPFL), Observatoire de Sauverny, 1290 Versoix, Switzerland\vspace*{-2pt} \\ 
\scriptsize $^{23}$ National Astronomy Observatories, Chinese Academy of Science, Beijing, 100012, P.R. China\vspace*{-2pt} \\ 
}

\date{Accepted XXX. Received YYY; in original form ZZZ}

\usepackage{hyperref}
\hypersetup{
    colorlinks=true,
    linkcolor=blue,
    citecolor=blue,
}

\begin{document}
\label{firstpage}
\pagerange{\pageref{firstpage}--\pageref{lastpage}}\pubyear{2020}
\maketitle

\begin{abstract}
We measure the clustering of quasars of the final data release (DR16) of eBOSS. The sample contains $343\,708$ quasars between redshifts $0.8\leq z\leq2.2$ over $4699\,\mathrm{deg}^2$. We calculate the Legendre multipoles (0,2,4) of the anisotropic power spectrum and perform a BAO and a Full-Shape (FS) analysis at the effective redshift $z{\rm eff}=1.480$. The errors include systematic errors that amount to 1/3 of the statistical error. The systematic errors comprise a modelling part studied using a blind N-Body mock challenge and observational effects studied with approximate mocks to account for various types of redshift smearing and fibre collisions.
For the BAO analysis, we measure the transverse comoving distance $\dmr{}=30.60\pm{0.90}$ and the Hubble distance $\dhr{}=13.34\pm{0.60}$. This agrees with the configuration space analysis, and the consensus yields: $\dmr{}=30.69\pm{0.80}$ and $\dhr{}=13.26\pm{0.55}$. 
In the FS analysis, we fit the power spectrum using a model based on Regularised Perturbation Theory, which includes Redshift Space Distortions and the Alcock-Paczynski effect. The results are $\dmr{}=30.68\pm{0.90}$ and $\dhr{}=13.52\pm{0.51}$ and we constrain the linear growth rate of structure $\fse{}=0.476\pm{0.047}$. Our results agree with the configuration space analysis. The consensus analysis of the eBOSS quasar sample yields: 
$\dmr{}=30.21\pm{0.79}$, $\dhr{}=3.23\pm{0.47}$ and $\fse{}=0.462\pm{0.045}$ and is consistent with a flat $\Lambda {\rm CDM}$ cosmological model using Planck results.

\end{abstract}



\section{Introduction}

Understanding the expansion history of the Universe is one of the crucial questions in cosmology. The latest results from the measurements of the angular temperature and polarisation fluctuations in the cosmic microwave background~\citep{Planck18} and the analysis of type Ia supernovae light curves~\citep{Pantheon} highly favors a Universe that can be described in the framework of General Relativity (GR) by a standard cosmological model, $\Lambda$CDM. In this model, the Universe is made of collisionless cold dark matter (CDM), baryons, photons, and neutrinos and of an unknown component, usually called `dark energy' which behaves as a fluid of negative pressure. In the $\Lambda$CDM context, a cosmological constant $\Lambda$ is inserted in the equation of general relativity to take account of the late-time acceleration of the expansion of the Universe.

In the last 15 years, this picture of the Universe has been shown to work remarkably well using the phenomenon of baryon acoustic oscillations (BAO) in the primordial plasma. BAO leave their imprint on the distribution of matter in the Universe as a characteristic separation scale between matter over-densities. This distance is found in the separation of gravitationally collapsed structures such as galaxies~\citep{Eisenstein+2005,Cole+2005,boss-dr12} and quasars~\citep{DR14-bao} and can be used as a `standard ruler' by large-scale surveys to measure the evolution of the expansion of the Universe at different epochs. 

As the effort to measure the BAO scale to increasingly better precision continues, large-scale surveys have started to provide valuable information on the linear growth rate of structure. This is of significant importance as it is a promising way to test GR~\citep{LinderCahn07}. 

The growth of structure is measured from coherent peculiar velocities that lead to Redshift Space Distortions (RSDs) along the line of sight~\citep{Kaiser87}. These distortions can be related to $f(z)\sigma_8(z)$, where $\sigma_8(z)$ is the normalization of the linear power spectrum on scales of $8\Mpch$ at redshift $z$ and $f$ is the linear growth rate of structure. Anisotropies in the clustering signal may also appear because the cosmology assumed to convert redshift to distance is different from the true cosmology. This is known as the Alcock-Paczynski effect~\citep{AP} and is key to constraining the cosmological expansion history.

In this paper, we present and analyse the power spectrum of the complete quasar sample of the extended Baryon Oscillation Spectroscopic Survey~\citep[eBOSS;][]{ebossoverview} which is part of the SDSS-IV program~\citep{sdss4}. The observations were made at the $2.5\;\mathrm{m}$ Sloan Foundation Telescope~\citep{gunn06} at the Apache Point Observatory (New Mexico, USA) using the two-arm optical spectrograph of BOSS~\citep{bossspectrometer}. This study is part of a coordinated release of the final eBOSS measurements of BAO and RSD in the clustering of luminous red galaxies \citep[LRG ($0.6<z<1.0$);][]{Bautista+20,Gilmarin+20}, emission line galaxies \citep[ELG ($0.6<z<1.1$);][]{Raichoor+20,Tamone+20,DeMattia+20}, and quasars \citep[($0.8<z<2.2$);][]{Hou+20}\footnote{A summary of all SDSS BAO and RSD measurements with accompanying legacy
figures can be found here: 
https://www.sdss.org/science/final-bao-and-rsd-measurements/.  The full
cosmological interpretation of these measurements can be found here: 
https://www.sdss.org/science/cosmology-results-from-eboss/.}. At the highest redshifts ($z>2.1$), the coordinated release of final eBOSS measurements includes measurements of BAO in the Lyman-$\alpha$ forest \citep{Dumasdesbourboux+20}. The cosmological interpretation of these results in combination with the final BOSS results and other probes is found in \citet{Eva}.

Due to their high intrinsic luminosity, quasars can be used as tracers of the large scale structure at high redshifts~\citep{Cro09,Myers+07,RossN+09,Shen+09,white12,Karagiannis+14,Eftekharzadeh+15,Laurent+16}. The Data Release 14 of the first two years of eBOSS data~\citep{Ata+18,Hector,Hou+18,Zarrouk+18} demonstrated how well quasars are suited for cosmological clustering analyses and currently provide the most precise clustering information on large scales in the redshift range $0.8<z<2.2$. With the Data Release 16, the number of quasars is more than doubled. We present the measurement of the redshift space power spectrum with the first three even Legendre multipoles. We perform both a standard `BAO-only' analysis where we focus on the BAO features of the power spectrum and a `Full-Shape' RSD analysis using the TNS model~\citep{TNS10}. The BAO-only analysis allows us to constrain the Hubble distance, $D_{\mathrm{H}}(z)/r_{\rm drag}$, and the transverse comoving distance, $D_{\mathrm{M}}(z)/r_{\rm drag}$. In addition, we also constrain these two quantities together with the linear growth rate of structure, $f(z)\sigma_8(z)$, using the `Full-Shape' RSD analysis.

The analysis presented in this paper uses the complete five years of the eBOSS sample and is accompanied by several companion papers. The clustering catalogues used in this analysis are described in~\citet{DR16} and specific information relevant to the complete DR16Q quasar catalogue is given in~\citet{DR16Q}. The quasar mock challenge upon which the model of the power spectrum is tested is described in~\citet{Smith+20}. Approximate mocks used for determining the covariance matrix and testing observational systematic effects are described in~\citet{ZhaoC+20}. The analysis of the quasar sample in configuration space is presented in~\citet{Hou+20} and a consensus analysis of the work presented here is common to both articles. Cosmological implications of the measured quasar clustering properties are discussed in~\citet{Eva}.

This paper is structured as follows. In Section~\ref{sec:Data}, we give an overview of the quasar sample, the estimator of the power spectrum and the set of mocks that we used for the estimation of the covariance and the assessment of the systematic errors. In Section~\ref{sec:BAO}, we discuss the measurement of the BAO scales. In Section~\ref{sec:FS}, we present the Full Shape RSD analysis and describe the systematic errors that affect the measurement. Our final result, and the consensus analysis performed in our companion paper on the 2-point correlation function analysis, are presented in Section~\ref{sec:conclusion}.

\section{Catalogues, Methods and Mocks}

\label{sec:Data} 
In this section, we describe the DR16 QSO catalogue and the method used calculate the power spectrum. We describe the EZmocks used for computing the covariance and testing systematic effects, and the mocks from the OuterRim N-body simulation used for testing the RSD and BAO models.
 
\subsection{Data catalogues}
\label{sec:catalogues}
The creation of the catalogues is fully described in~\citet{DR16}, and we summarize in this section the information relevant to the eBOSS DR16 quasar sample. Quasar targets were selected according to the procedure presented in~\citet{Myers+15}. It is based on the SDSS-I-II-III optical imaging data in the {\it ugriz} photometric pass bands~\citep{Fukugita+96} and on the Wide-field Infrared Survey Explorer~\citep{Wright+10}. The selection algorithm uses the `extreme deconvolution' technique~\citep{Bovy+11} to select a homogeneous quasar targets sample over the footprint. The set of selected targets are then processed by the tiling algorithm~\citep{Blanton+2003} to assign spectrograph fibres to a maximum number of targets. The most important feature of this algorithm is that it assigns a fibre to each target of a `decollided set' for which at least one object in a collision group (separated by less than 62", the physical size of the fibre holder) has a fibre assigned. The weight of the unobserved targets is transferred to the other members of the collision group that received a fibre. Each fibred object then has a close pair weight, $w_{\rm cp}$, that allows the large-scale angular clustering to be recovered. In Section~\ref{sec:systs}, we present the impact of fibre collisions on our measurements, and the techniques used to mitigate it. The probability to reliably measure the redshift depends on the signal to noise ratio of the spectrum and correlates spatially with both the coordinate in the focal plane, and with the location of the spectrum on the CCD of the spectrograph. In practice, this probability is best measured as a function of the number {\it ID} of the spectrograph fibre and an additional correction for the overall spectrograph signal to noise ratio in the i-band is applied. The inverse of this probability is used as a weight, $w_{\rm noz}$, to account for missing objects due to redshift failures. Then, a weight, $w_{\rm sys}$, is used to account for the fact that the density of targets depends on the imaging conditions~\citep{RossA+12, DR14-bao, Bautista+18}. It is determined from a multivariate linear regression of the angular density of targets with respect to 4 different photometric quantities such as extinction (E[B-V]) and depth in the g-band corrected for extinction (See~\citealt{DR16} for details on the full procedure). Finally, following~\citet{Feldman+94}, we apply a weight, $w_{\rm FKP}=(1+\bar{n}P_0)^{-1}$, to minimize the estimator variance. This weight depends on the weighted number density of the sample, $\bar{n}$, and on $P_0$ the power spectrum at a typical scale. In the present sample, we use a value of $P_0=6000\Mpchc{}$, that is the value of the power spectrum at $k=0.14\hMpc{}$ that is in the middle of the range of scales under consideration and where the BAO signal is the most prominent in our analysis.
The total weight, $w_{\rm tot}$ that is applied to each object for the power spectrum measurement is given by the product of these weights,
\begin{equation}
w_{\rm  q}=w_{\rm cp}\cdot w_{\rm noz}\cdot w_{\rm sys}\cdot w_{\rm FKP}.
\label{eq:weights-q}
\end{equation}
The clustering of our sample is calculated by comparing our data to a random catalogue that has no physical clustering, but is designed to have the same angular and radial selection function as the data. The random catalogue is obtained by randomly drawing objects in the angular coverage of the footprint of the data, and the redshift of each object is randomly drawn from the redshift distribution of the data. Contrary to what was done in the DR14 analysis~\citep{Hector}, the random catalogues are not downsampled by the completeness but are weighted by the completeness ($C_{\rm eBOSS}$) of the sector to match the data. The weights of the objects in the random catalogue are then
\begin{equation}
w_{\rm  r}=C_{\rm eBOSS}\cdot w_{\rm FKP}.
\label{eq:weights-r}
\end{equation}
The weighted area and the number for quasars used for the present analysis are given in Table~\ref{tab:area}). The number of objects in the random catalogue is taken to be 50 times the number of objects in the data catalogue.

\begin{table}
	\centering
	\caption{Effective area and number of quasars used in the clustering analyses.}
	\label{tab:area}
	\begin{tabular}{lccc} 
		\hline
		cap & NGC & SGC & total \\
		\hline
		Weighted area (deg$^2$)     & 2860    & 1839    & 4699 \\
		Quasars used $0.8<z<2.2$  & $218\,209$ & $125\,499$ & $343\,708$ \\
		\hline
	\end{tabular}
\end{table}

\subsection{Estimation of the power spectrum and extraction of cosmological parameters}

We study the quasar survey through its power spectrum;
\begin{equation}
    P(\bf k) = (2\pi)^3 \delta^D({\bf k} + {\bf k^\prime})\langle \delta(\bf k)\delta(\bf k^\prime) \rangle
\end{equation}
where $\delta({\bf k})$ is the density perturbation field and $\delta^D$ is the Dirac distribution.

\begin{figure}
	\includegraphics[width=\columnwidth]{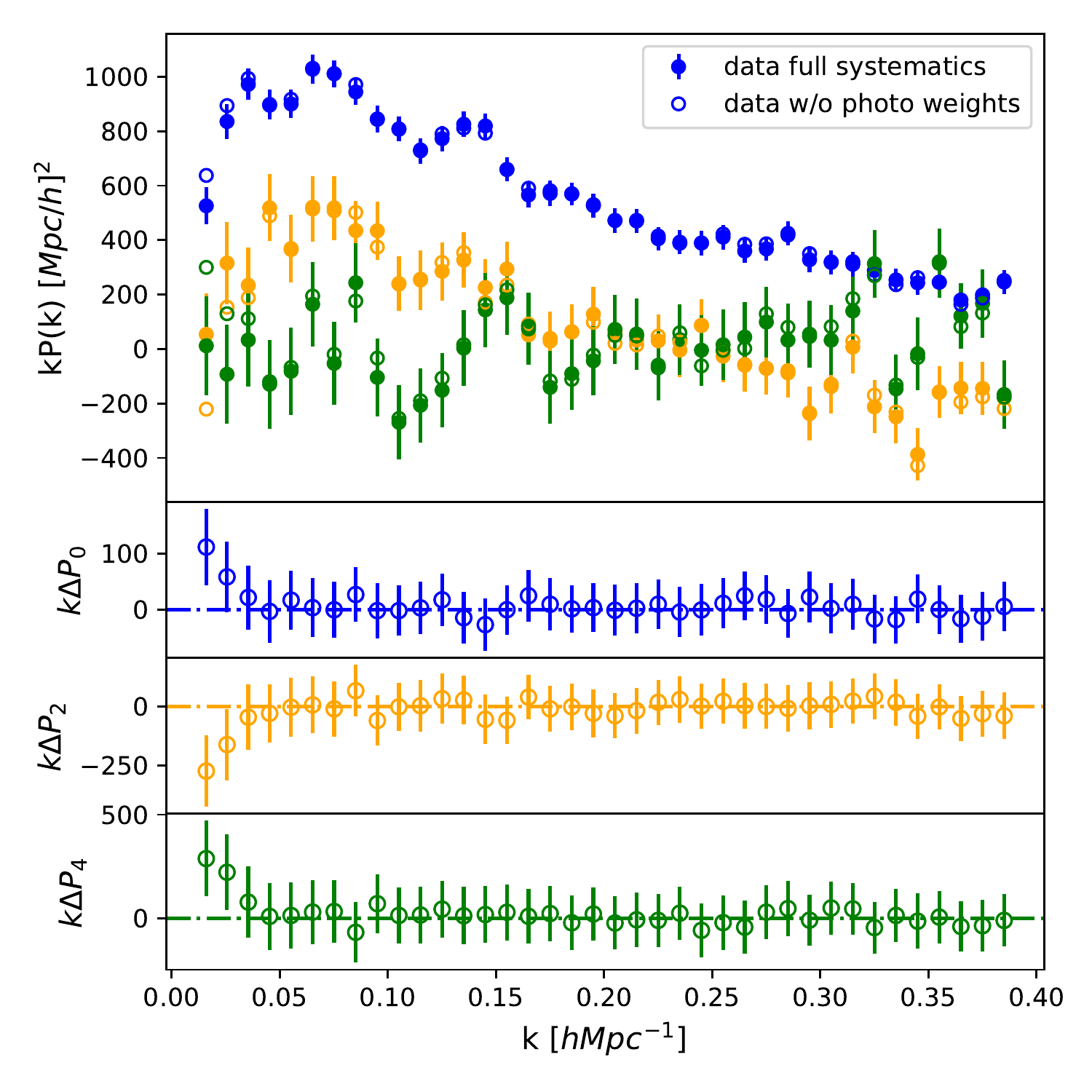}
	\includegraphics[width=\columnwidth]{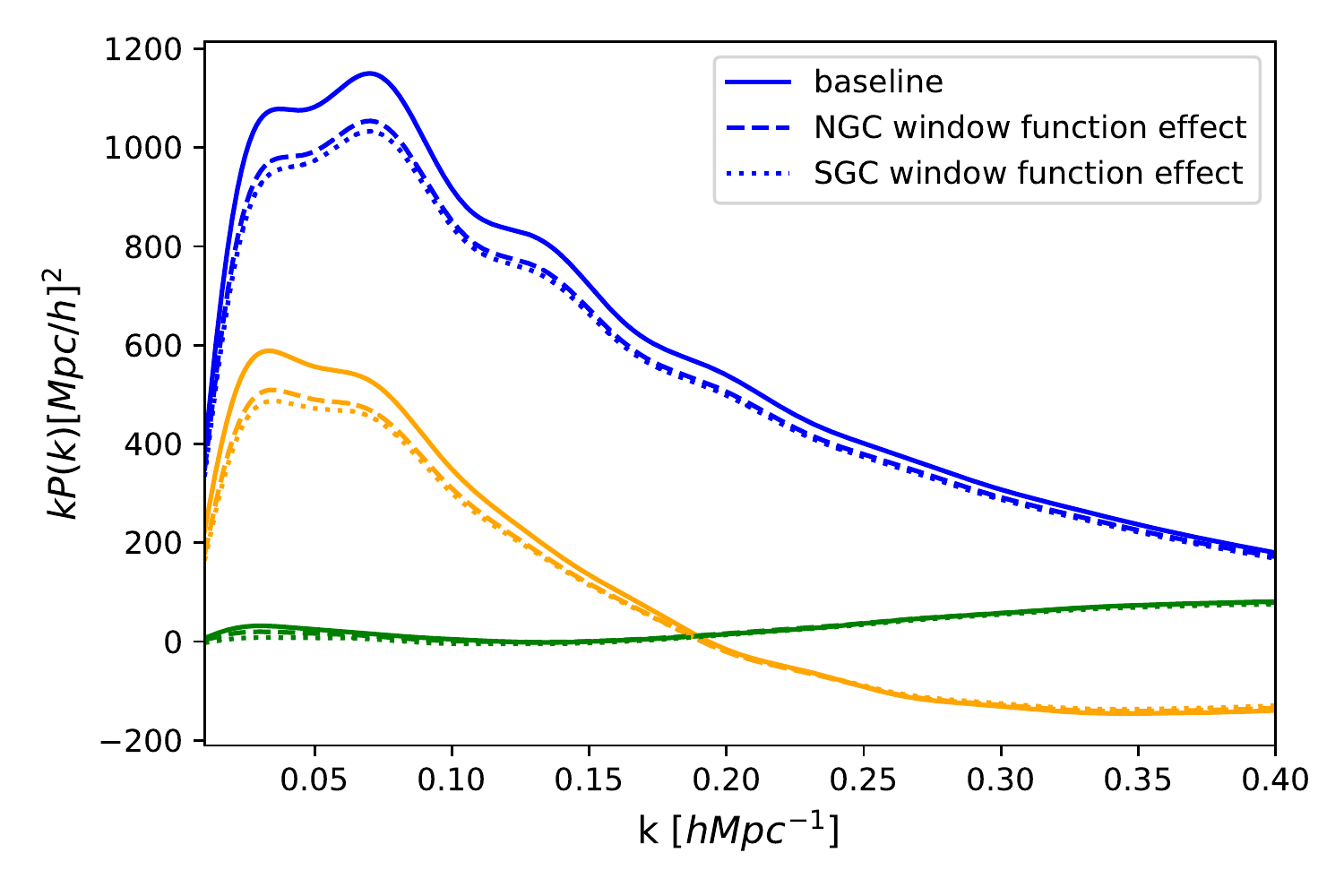}
	\caption{Top panel: Power spectrum of the SGC data with all weights applied (solid circle) or without the photometric weight (open circle); the effect on the NGC (not shown here) is smaller. Represented are the multipoles of the power spectrum:  monopole (blue), quadrupole (red), and hexadecapole (green). 
	Lower panel: Impact of the NGC (dashed line), and SGC (dotted line) window function on the power spectrum multipoles of a baseline power spectrum (solid line, same color scheme as in the top panel) }
    \label{fig:window-and-syst}
\end{figure}
For the calculation of the power spectrum, we use the method of~\citet{Yamamoto+2006}. First, we construct a weighted object over-density field, $F({\bf r})$~\citep{Feldman+94}, at each position ${\bf r}$,
\begin{equation}
F({\bf r})=n_q({\bf r})-\alpha n_{r}({\bf r}),
\label{eq:fourier_fkp_field}
\end{equation}
where $n_q$ is the number density field for the quasar catalogue and $n_{r}$ is the number density of objects in the random catalogue described in Section~\ref{sec:catalogues}. Both are including the weights described in Eqs.~\ref{eq:weights-q} and~\ref{eq:weights-r}. The random field is scaled by $\alpha=\frac{\sum_i w_{q,i}}{\sum_i w_{r,i}}$ to ensure that the mean weighted densities of data and random are the same. Then, following~\citet{Bianchi_2015} the power spectrum multipoles are given by: 
\begin{equation}
P_\ell(k)=\frac{2\ell+1}{I}\int\frac{d\Omega_k}{4\pi}\left[F_0({\bf k})F_\ell({-\bf k})-P^{\rm noise}_0(k)\right],
\label{eq:yamamoto}
\end{equation}
where I is the normalization of the power spectrum, defined as
\begin{equation}
I=\alpha \sum_{i=1}^{N_r} w_{r,i}n_{q,i},
\label{eq:norm_PS}
\end{equation}
and ${\cal L}_\ell$ is the $\ell$th order Legendre polynomial. $\Omega_k$ is the solid angle in Fourier space, and $F_{\ell}({\bf k})$ is defined as
\begin{equation}
F_{\ell}({\bf k}) = \int d^{3} r F({\bf r}) {\cal L}_\ell(\hat{\bf k} \cdot \hat{\bf r}) e^{i {\bf k} \cdot {\bf r}}.
\end{equation}
The shot-noise component, $P^{\rm noise}_\ell(k)$, is expressed as
\begin{equation}
P^{\rm noise}_0(k)=\sum_{i}w^2_{q,i} + \alpha^2\sum_{i}w^2_{r,i}.
\label{eq:shot_noise}
\end{equation}

In practice, we use the \texttt{nbodykit} package~\citep{nbodykit} to calculate the power spectrum multipoles using the method of~\citet{Hand+2017}. First, the weighted density field is mapped onto a cubic grid using the Triangular Shaped Cloud (TSC) interpolation method. Each cap is enclosed in a box of dimensions $L_{box}=[3100,6500,2700]\Mpch$. The cell size is chosen to be $7\Mpch$, yielding a Nyquist frequency of $k_{\rm Nyq}=0.449\hMpc$ well above the maximum wave number of our analysis ($k_{\rm max}=0.3\hMpc$). 
Then, the $F_{\ell}({\bf k})$ term can be computed with a Fast Fourier Transform, and the interlacing technique is used to reduce the effect of aliasing~\citep{Sefusatti_2016}.  
In the top panel of Fig.~\ref{fig:window-and-syst}, we show the impact of photometric weights in the calculation of the power spectrum for the South Galactic Cap, which is known to be the most affected by photometric systematics, as demonstrated in~\citet{Zarrouk+18}. We observe that the correction brought by the photometric weight changes the multipoles on scales $k<0.05\hMpc$. In Sect.~\ref{sec:syst-rsd} we use the approximate mocks to evaluate the impact of applying and correcting for systematic effects.

The observed density of quasars is measured according to a survey-specific selection function that must be folded into the model. Following the method described in~\citet{Beutler_2016}, the selection function multipoles are determined in configuration space from the weighted number of pairs in the `random' catalogue as 
\begin{equation}
W_\ell(r) \propto \sum_{\mu}\sum_{x_1}\sum_{x_2}RR(r,\mu)L_\ell(\mu),
\label{eq:window}
 \end{equation}
with $RR(r,\mu)$ being the random-random pair counts in bins of separation $r$ and $\mu$, where $\mu$ is defined as the cosine of the angle between the line of sight and the line connecting the two objects of a pair. As proposed in~\cite{DeMattia+2019}, the window function is normalised by the quantity $I$ (Eq.~\ref{eq:norm_PS}), in order to be consistent with the estimator of the power spectrum, which ensures that the model fitting is independent of the choice of normalization. 
The effect of the window function on the power spectrum multipoles, using the power spectrum model described in Section~\ref{sec:model}, is shown in the bottom panel of Fig.~\ref{fig:window-and-syst}. Small differences between the two galactic caps are observed and arise from the difference in angular size and shape of their respective footprints (the SGC footprint is indeed smaller and patchier).

As a single line of sight is used for each galaxy pair, wide-angle effects may arise and are taken into account using the formalism described in~\citet{Beutler_2019} which consists of expanding the survey window function in $s/d$, the ratio of pair separation to the comoving distance from the observer.

In the analyses performed hereafter, we assume a fiducial cosmology to convert redshifts to distances along the line of sight, and to calculate a power spectrum template:
\begin{equation}
\begin{split}
h &= 0.676, \,\, \Omega_{m} = 0.31, \,\,  \Omega_{\Lambda} = 0.69,\\
\Omega_{b}h^{2} &= 0.022, \,\, \sigma_{8} = 0.80. \qquad
\end{split}
\label{eq:fiducial-cosmology}
\end{equation}
To fit to the data, we allow for variations of the dilation scales along ($\alpha_\parallel$) and perpendicular ($\alpha_\perp$) to the line of sight, following the test proposed by~\citet{AP}. Changing the dilation scales in configuration space is equivalent to setting $k'_{\parallel}=k_{\parallel}/\alpha_{\parallel}$ and $k'_{\perp}=k_{\perp}/\alpha_{\perp}$. This change of scales corresponds to the transformation from $(k,\mu)$ to $(k',\mu')$ where
\begin{equation}
\begin{split}
k&'=\frac{k}{\alpha_{\perp}}\left[1+\mu^2\left(\frac{1}{F^2}-1\right)\right]^{1/2}\\
\mu'&=\frac{\mu}{F\left[1+\mu^2\left(\frac{1}{F^2}-1\right)\right]^{-1/2}},\\
\end{split}
\end{equation}
where $F=\alpha_{\parallel}/\alpha_{\perp}$. 

\begin{figure}
	\includegraphics[width=\columnwidth]{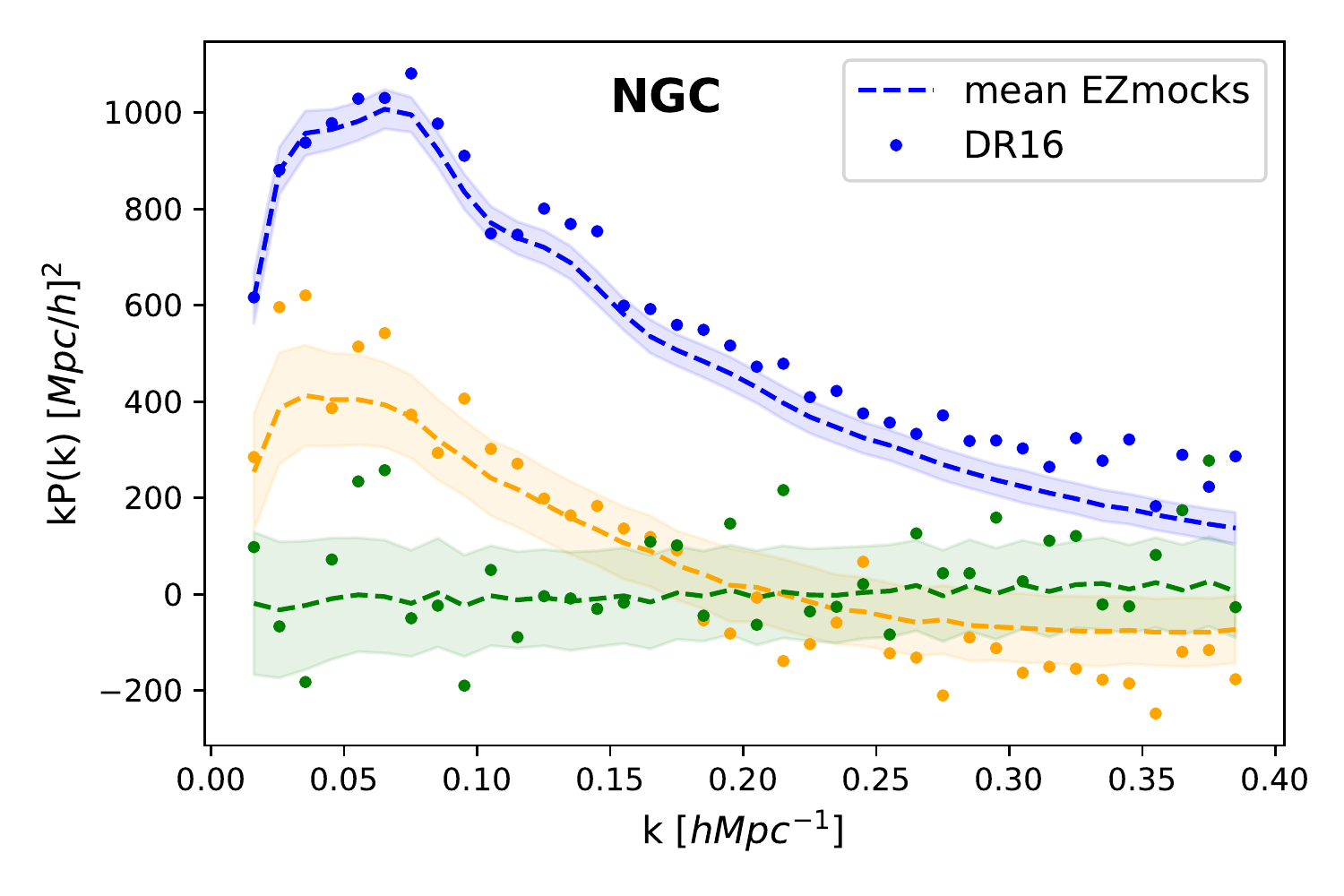}
	\includegraphics[width=\columnwidth]{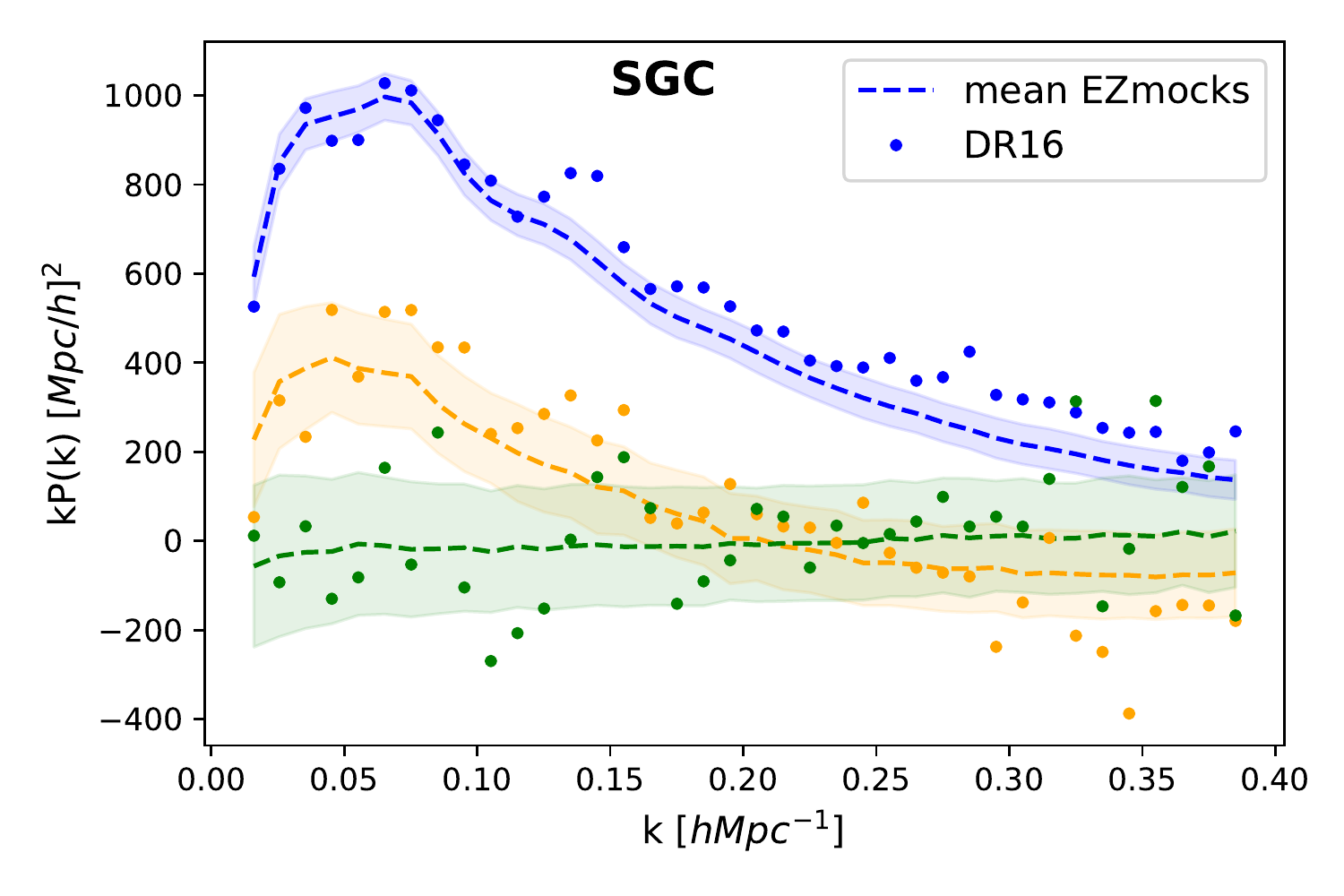}
    \caption{Multipoles of the power spectrum measured with the DR16 eBOSS quasar sample (dots) compared to the EZmocks (dashed line). The standard deviation of the mocks is indicated by the shaded area. The NGC and SGC are shown in the top and bottom panels, respectively. The monopole is shown in blue and the shot noise contribution is subtracted, the quadrupole in red and the hexadecapole in green.}
    \label{fig:ps-data-ez}
\end{figure}

Provided the constraints on $\alpha_{\parallel}$ and $\alpha_\perp$ only come from the BAO location, these quantities are related to the Hubble distance $D_{\rm H}\equiv c/H(z)$ where $H(z)$ is Hubble expansion, and to the transverse diameter distance $D_{\rm M}(z)$ independently of the choice of fiducial cosmology through the following expressions: 
\begin{equation}
\label{eq:cosmo_dilationscale}
\alpha_{\parallel}=\frac{D_{\rm H}(z)/r_{\rm drag}}{D_{\rm H}^{\rm fid}(z)/r_{\rm drag}^{\rm fid}},\qquad \alpha_{\perp}=\frac{D_{\rm M}(z)/r_{\rm drag}}{D_{\rm M}^{\rm fid}(z)/r_{\rm drag}^{\rm fid}},
\end{equation}
where the quantities with the superscript `fid' are determined within the fiducial cosmology and $r_{\rm drag}$ is the comoving sound horizon at $z=z_{\rm drag}$, where $z_{\rm drag}$ is the redshift at which the baryon-drag optical depth equals unity~\citep{HuSugiyama1995}.
In the case of spherically averaged clustering measurements, the analysis is sensitive to an isotropic dilation scale, $\alpha_{\mathrm{iso}}$, that can be expressed as
\begin{equation}
\alpha_{\mathrm{iso}}=\frac{D_{\rm V}(z)/r_{\rm drag}}{D_{\rm V}^{\rm fid}(z)/r_{\rm drag}^{\rm fid}},
\end{equation}
where $D_{\rm V}(z)$ corresponds to the spherically averaged BAO distance, that we can link to the other cosmological parameters through the relation~\citep{Eisenstein2005:astro-ph/0501171v1,Ross2015:1501.05571v2}:
\begin{equation}
D_{\rm V}(z)=\left[czD_{\rm H}(z)D_{\rm M}(z)\right]^{1/3}.
\end{equation}

The linear growth rate of structures, $f$, is determined from a 'Full-Shape' fit of the  power spectrum multipoles. In practice, the non linear power spectrum is calculated assuming a linear power spectrum of known normalisation which is proportional to $\sigma_8$, the amplitude of matter perturbations below scales of $8\Mpch$. In linear theory, $f$ and $\sigma_8$ are completely degenerate~\citep{Percival+2009}, and hence, our measurement is sensitive to the product, $f(z)\sigma_8(z)$, at the effective redshift of the survey.

The definition of the effective redshift adopted in this analysis is
\begin{equation}
z_{\rm eff}=\frac{\sum_i \sum_j w_i w_j (z_i+z_j)/2}{\sum_i \sum_j w_i w_j }=1.480\pm0.001,
\end{equation}
where the sums are restricted to separations between $25$ and $120\Mpch$, as this corresponds to the scales over which the fits are performed in configuration space. The error quoted is the statistical error on the unweighted redshift distribution.

\subsubsection{Parameter estimation}

For both the BAO analysis (Section~\ref{sec:BAO}) and the full shape RSD analysis (Section~\ref{sec:FS}),
the best-fit model parameter values are obtained by finding the point in parameter space that maximizes the likelihood function,
\begin{equation}
    L \propto e^{-\chi^2/2},
   \label{eq:likelihood}
\end{equation}
with the quantity $\chi^2$ defined by
\begin{equation}
    \chi^2=(P_{\rm meas}-P_{\rm model})^t W(P_{\rm meas}-P_{\rm model}).
     \label{eq:chi2}
\end{equation}
$P_{\rm meas}$ is the data vector of the power spectrum multipoles measurement in intervals of wave number $k$, and $P_{\rm model}$ is the corresponding vector for the model. In this expression, $W$ is the inverse of the covariance matrix computed from EZmocks, described in Section~\ref{sec:EZmock}, which we correct for the finite number of mocks following \citet{Hartlap+2007}:
\begin{equation}
W = \frac{N-n-2}{N-1}C^{-1},
\label{eq:hartlap}
\end{equation}
where $N$ is the total number of mocks, and $n$ is the number of data points.
Furthermore, due to the uncertainty in the covariance matrix, a correction on the variance of the fit parameters is applied, following~\citet{Percival+2014}.
The total $\chi^2$ combines the NGC and SGC with a common set of $\alpha_\perp$, $\alpha_\parallel$ and $f\sigma_8$ parameters (or just $\alpha_\perp$ and $\alpha_\parallel$ for the BAO-only analysis). The nuisance and bias parameters are allowed to differ for the two galactic caps. We determine the covariance matrices independently in each cap assuming that there is no cross-correlation between the two caps. Therefore, the number of points $n$ to be used in Eq.~\ref{eq:hartlap} corresponds to the number of data points in a single cap (i.e. 84 over the full $k$-range, including all even multipoles up to $\ell=4$).

We use, for the final result of the Full Shape RSD analysis, the likelihood function defined in Eq.~\ref{eq:likelihood} to produce Monte Carlo Markov Chains with the \texttt{emcee} package~\citep{emcee}. We check the convergence of the chains with the Gelman-Rubin convergence test requiring $R<0.01$. The $\chi^2$ minimisation is performed using the MINUIT\footnote{MINUIT, Function Minimization and Error analysis, James F., CERN 1994} program libraries. In this case, parameter errors are determined from finding the $\Delta\chi^2=1$ abscissa along the 1D $\chi^2$ profile for each parameter. After ensuring the errors of both techniques are compatible, we apply this frequentist method for all the results of this paper concerning mocks as well as the various tests done on the data as it is much faster (usually, the fit outperforms the MCMC running by a factor of ~1000 in terms of CPU time).

\subsection{Mocks}

We present, in the following, the two sets of mocks that we use; the approximate EZmocks are used to estimate the observational systematic errors and the computation of the covariance matrix, and the mocks created from N-body simulations are used to derive the modelling systematic errors.

\subsubsection{EZmocks}
\label{sec:EZmock}

To determine the covariance matrix to be used in the fits to the data and to test our analysis pipeline, we use a set of 1000 approximate light-cone mock catalogues that are extensively described in~\citet{ZhaoC+20}. These mocks, dubbed EZmocks, are made using the framework of the Effective Zel'dovich approximation developed in~\citet{EZmocks}. The number of mocks (i.e. 1000) is chosen such that the correction given in Eq.~\ref{eq:hartlap} is below 10\%. The simulation box is a cube with a comoving side length of $5\Gpch$. The $\Lambda$CDM cosmology assumed when creating the EZmocks has the following parameters:
\begin{equation}
\begin{split}
h &= 0.6777, \,\, \Omega_{m} = 0.307115, \,\,  \Omega_{\Lambda} = 0.692885,\\
\Omega_{b} &= 0.048206, \,\, \sigma_{8} = 0.8225, \,\,  n_s=0.9611\qquad
\end{split}
\label{eq:cosmo_EZ}
\end{equation}
For the quasars, the light-cone mocks are built from 7 simulation snapshots, and the clustering signal in the mocks is tuned to match the one of the final DR16 quasar catalogues (see Fig.~\ref{fig:ps-data-ez}). At the scale of the second BAO wiggle, $k\sim0.14 \hMpc$, the data is in excess of power compared to the EZmocks. That is due to a stronger BAO signal in the data. The EZmocks are mocks generated with a fast-technique, which is a good approximation of an N-body simulation mock at large scales, but may fail to reproduce the non linear gravity interactions at small scales accurately. This might explains the mild lack of power on the monopole of the EZmocks corrected for shot-noise for both caps at $k>0.25 \hMpc$. 

In practice, it is the monopole not corrected for shot noise of the EZmocks that is used in the determination of the covariance matrix. For this quantity, we see that the EZmocks tend to overestimate the monopole by 3 percent and this will result in a conservative overestimation of the errors on the fit parameters by the same amount.

Furthermore, the power spectrum monopole for the quasar sample is increasingly dominated by shot-noise as $k$ increases. As the weighted number of objects in the mocks are matched to the data by construction, the shot-noise terms are the same and the impact of the lack of power at small scales is reduced. Quantifying the residual impact this has on the measurement of the cosmological parameters is addressed in the studies of systematic effects in the next Sections.

Mocks are also used to estimate the impact of systematic effects present in the data. To do so, the approximate EZmocks are modified to reflect the effects induced by observational conditions. Firstly, mock `data' catalogues are created by taking mock quasars in the redshift range $0.75<z<2.25$. The mock catalogues are downsampled by an amount that allows to match the radial selection function of the data at the end of the procedure. Then, contaminants (stars, galaxies, and `legacy' quasars), that were known before the quasar survey and that fulfilled the quasar target selection conditions, are added to this catalogue. The fibre assignment algorithms \citep[based on \texttt{nbodykit};][]{nbodykit} is run on this set of targets using the plate geometry of the DR16 data. As in the data, objects that could not receive a fibre are treated by up-weighting the objects in the collision group that did receive a fibre. The effects of the imaging conditions are modeled by varying the number of targets according to the weight maps measured in the data. We use the spectroscopic success rate as a function of the identification number of the spectrograph fibre measured in the data as well as the plate signal to noise ratio to randomly remove objects. As a consequence, the objects in the mock catalogues receive a weight to cope for the spectroscopic success rate variations, as was described for the data catalogues. 

\subsubsection{OuterRim mocks and the quasar mock challenge}
\label{sec:ORmock}
To test the model adopted for the power spectrum we compare it against accurate mock catalogues of known cosmological parameters.
For this purpose, we have performed a quasar mock challenge, which is described in details in~\citet{Smith+20}. We refer the reader to this article, and we briefly summarise the content and give the results here. The mocks are built from one snapshot ($z=1.433$) of the OuterRim N-body simulations~\citep{OuterRim} that was available at the beginning of this work. Non-linear effects are mildly increasing with time, therefore the tests of the model are performed conservatively since the effective redshift of the DR16 quasar sample corresponds to earlier times.
The OuterRim simulation covers a cubic box of length $L=3\Gpch$ with periodic boundary conditions and uses $10240^3$ dark matter particles. The mass of each particle in the simulation is hence $m_p = 1.82 \times 10^9\,{\rm M_\odot}h^{-1}$ and dark matter haloes are detected using a Friends-of-Friends algorithm. Initial conditions are calculated at $z = 200$ in the Zel'dovich approximation using the cosmological parameters:
\begin{equation}
\begin{split}
h &= 0.71, \,\, \Omega_{\mathrm{cdm}}h^{2} = 0.1109, \,\,  \Omega_{\Lambda} = 0.735,\\
\Omega_{b}h^{2} &= 0.02258, \,\, \sigma_{8} = 0.80, \,\,  n_s=0.963,\qquad
\end{split}
\label{eq:cosmo_OR}
\end{equation}
which are consistent with WMAP7~\citep{WMAP7}.
To populate dark matter haloes with quasars, we consider various Halo Occupation Distribution (HOD) models. An HOD describes the probability that a halo of a given mass contains central or satellite quasars. Central quasars are placed at the centre of the halo, while satellites are placed according either according to a NFW profile~\citep{NFW} or by drawing a particle from a $1\%$ subsample of the particles belonging to the halo. The list of HOD parameters can be found in~\citet{Smith+20}. For each HOD model, we generate 100 random realisations. We also reproduce the effect of redshift uncertainties, which are due to broad emission lines in the quasar spectra, and have been shown to be the leading effect which impacts the recovery of cosmological parameters \citep{Zarrouk+18}.
Three redshift smearing prescriptions were applied: no redshift smearing, Gaussian smearing and a realistic smearing case with non-Gaussian tails. A fourth set of mocks has been created starting from the realistic smearing case and assigning $1.5\%$ of objects with a catastrophic redshift. We use these `non-blind' mocks to tune the functional form of the damping term that enters our model of the power spectrum (See Section~\ref{sec:model}), which plays a leading role at small scales.

In a second stage,~\cite{Smith+20} implemented the rescaling technique described in~\citet{Mead+2014} \citep[itself based on the work of][]{Angulo+2010} to create `blind' mocks, whose the true cosmology was only revealed at the end of the analysis. For this part of the mock challenge, two snapshots ($z=1.376$ and $z=1.494$) of the OuterRim N-body simulations have been rescaled to eight different cosmologies at $z=1.433$ and for three types of HOD. No modifications to the models were undertaken after the cosmologies were un-blinded.

\section{BAO analysis}
\label{sec:BAO}

In this section, we present the BAO-only analysis by first explaining the model, and then we present the results and the systematic tests performed.

\subsection{Model}
\label{sec:BAOmodel}

The modelling of the power spectrum for extracting the dilation scales parallel and perpendicular to the line of sight is based on the method described in~\citet{Kirkby+13}. In this method, the 2-point correlation function, which is the Fourier transform of the power spectrum, is decomposed  into a `smooth' term that characterises the broadband shape and into a `peak' term representing the BAO wiggle part of the power spectrum. In the fiducial analysis, the smooth term is not affected by the dilation scaling but in the analysis named `smooth-term coupling' we apply the dilation to both terms. The Fourier transform of these two terms, $P_{\rm sm}(k,\mu)$ and  $P_{\rm peak}(k,\mu)$, are used to model the measured power spectrum through:
\begin{equation}
    P(k,\mu)=\left[P_{\rm sm}(k,\mu)+P_{\rm peak}(k,\mu)e^{-\Sigma_\mathrm{nl}k^2}\right]\frac{b^2(1+\beta\mu^2)^2}{1+\frac{(k\mu\Sigma_s)^2}{2}},
\end{equation}
where $\Sigma_\mathrm{nl}$ is an anisotropic damping term of the BAO feature, $b$ represents the linear bias of the quasars, $\beta=f/b$ is linear enhancement of the power spectrum due to redshift space distortions and $\Sigma_s$ is another damping term accounting for random velocities at small scales and redshift uncertainties.
Given the low density of the quasar sample, the BAO reconstruction technique cannot be applied successfully as the matter field is not accurately sampled. 

The power spectrum is then decomposed into Legendre multipoles ($\ell=0,2,4$), which are Fourier transformed to obtain the corresponding correlation functions. The window function is applied on the correlation function multipoles using the method presented in~\citet{Beutler_2016}, and involves multipoles of the window function determined up to the $\ell=8$ order. A final Fourier transform is applied to the correlation function multipoles to obtain the window function convolved power spectrum multipoles.

Furthermore, to fit the broadband part of the PS, we add 3 polynomial terms to each multipole,
\begin{equation}
P^f_\ell(k) = P_\ell(k)+\frac{a_{0,\ell}}{k}+a_{1,\ell}+a_{2,\ell}\cdot k,
\end{equation}
where the $a_{x,\ell}$ are allowed to vary in the fitting procedure and are different for the two galactic caps. 

The anisotropic non linear damping of the BAO, $\Sigma_{nl}$, is modeled as 
\begin{equation}
\Sigma_{nl}=(1-\mu^2)\Sigma_{\perp}^2/2+\mu^2\Sigma_{\parallel}^2/2.
\end{equation}
The values of the damping parameters $\Sigma_{\parallel}$, $\Sigma_{\perp}$ and $\Sigma_s$ are obtained from fitting the non-blind N-body mocks, described in Section~\ref{sec:ORmock}. Firstly, we used the set of mocks without redshift smearing effects to determine $\Sigma_{\parallel}$ and $\Sigma_{\perp}$. Then, we fix $\Sigma_{\parallel}$ and $\Sigma_{\perp}$ to the obtained values and fit the mocks with realistic smearing to determine $\Sigma_s$. This procedure yields (in units of $\Mpch$):
\begin{equation}
    \Sigma_{\parallel}=8.5\pm0.6\qquad \Sigma_{\perp}=2.9\pm0.9 \qquad \Sigma_s=4.5\pm1.1.
\end{equation}

\begin{figure}
	\includegraphics[width=\columnwidth]{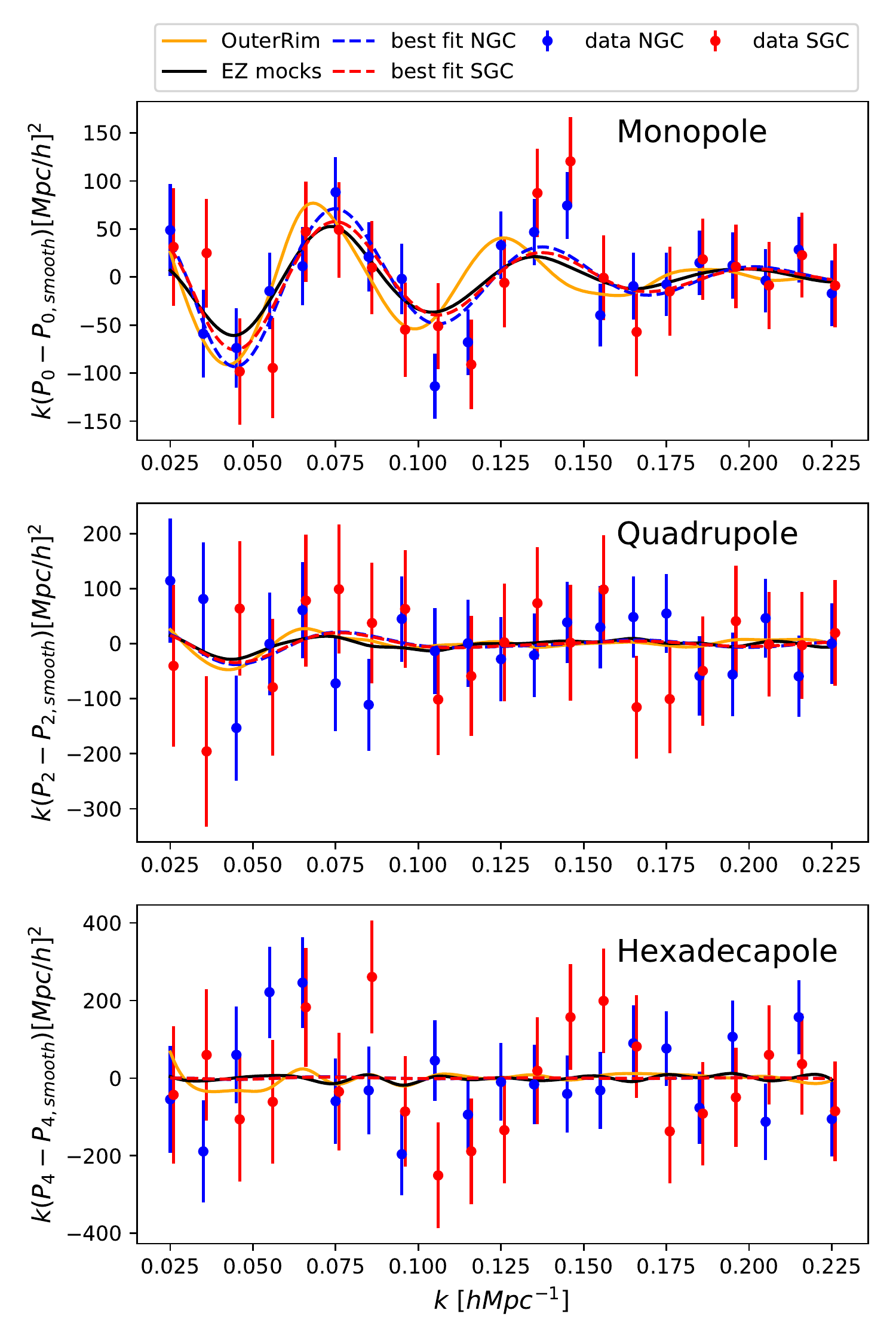}
    \caption{Comparison of the BAO wiggles in the power spectrum of the data and mocks. The dots represent the DR16 data, the dashed lines are the best-fit, the black line shows the mean of the NGC EZmocks. The green line shows the mean of one realisation of the OuterRim mock challenge (mock3) with realistic smearing. For the latter, the BAO feature appears shifted as a consequence of their intrinsic cosmology being different.}
    \label{fig:bao_wiggle}
\end{figure}

In Fig.~\ref{fig:bao_wiggle}, we present the BAO wiggle part of the power spectrum for the data, and for the average of the EZmocks. Also plotted is the average of the 100 realisations from one set of OuterRim non-blind mocks (note that it is at a different cosmology than the EZmocks). 
The data show a clear detection of the BAO for both galactic caps, and the amplitude of the oscillation is found to be larger than in the EZmocks. The amplitude of the oscillation in the EZmocks is itself smaller than the expectation in the case of the Zel'dovich approximation.

For the fiducial analysis, the fit is performed over wave numbers $k=[0.02, 0.23]\hMpc$ that cover the first three visible BAO oscillations. We use 22 free parameters: the 2 dilation scale parameters, 2 bias parameters (one for each cap) and 18 broadband terms (3 for each cap and each multipole), with intervals of allowed variations of $b=[0,10]$ and $\alpha_{\parallel,\perp}=[0.8,1.2]$ that are large enough that boundaries are never hit. The 3 damping terms are fixed to $[\Sigma_{\parallel}, \Sigma_{\perp} ,\Sigma_s]=[8,3,4]\Mpch$ since letting them free in the fit may result in an artificial improvement of the statistical precision~\citep{hinton2019barry}.

We use the mock challenge~\citep{Smith+20} to determine the systematic error originating from the modelling of the BAO feature in the power spectrum multipoles. The analysis is performed on all sets of OuterRim mocks in the non-blind and blind challenges, which includes the effects of different cosmologies, redshift error prescriptions, and different HOD models. The systematic errors are found to be
\begin{equation}
    \Delta(\alpha_\parallel)\bare{model}=0.0098 \quad  \Delta(\alpha_\perp)\bare{model}=0.0055,
\label{eq:syst-mc-bao}
\end{equation}
and are obtained from the standard deviation (w.r.t. the expected value) of the non-blind `realistic smearing + catastrophic redshifts' mocks added in quadrature with that of the blind mocks. 


\subsection{Results}

We perform the BAO-only analysis of the monopole, quadrupole and hexadecapole of the power spectrum measured from the QSO DR16 catalogue, as described in Section~\ref{sec:BAOmodel}. The covariance matrix is computed from the 1000 EZmocks. Fig.~\ref{fig:bao_wiggle} shows the best fit model (dashed lines) to the data (points) with NGC and SGC fitted simultaneously.
The results obtained from the BAO analysis in Fourier space are in terms of $\alpha$:
\begin{equation}
\alpha_\parallel=1.035 \pm 0.045 \qquad \alpha_\perp=1.017 \pm 0.029,
\label{eq:bao-results}
\end{equation}
where the errors are statistical only. In this section we estimate the systematic errors on this measurement using a series of tests performed on the EZmocks and we verify on the data that the choices made for the analysis do not bias the results.

\subsubsection{Systematic tests}
\label{sec:systs}

\begin{table}
	\centering
	\caption{Average $\alpha_\parallel$ and $\alpha_\perp$ values obtained for the 1000 EZmocks under different type of systematic effects and methods (first four rows) and results obtained for data with different model prescriptions and different damping values (bottom two rows). The 'no wf' line stands for the analysis performed without taking into account the window function correction, the 'coupled' line shows the impact of coupling the sideband of the model, the lines $w_{\rm noz}+w_{\rm sys}$ and fibre collisions show the effect of the redshift failures+photometric systematics or the collisions of fibres respectively.}
	\label{tab:bao-numbers}
	\begin{tabular}{llll} 
		\hline
		 tests on EZmocks  & $\alpha_\parallel$ & $\alpha_\perp$ & \\
		\hline
		Reference        & 0.9938$\pm$0.0027 & 0.9959$\pm$0.0019 & \\
        no wf            & $\Delta=0.0007(2)$ & $\Delta=0.0011(1)$ &\\
        coupled          & $\Delta=0.0031(11)$ & $\Delta=0.0021(6)$& \\
        no wf, coupled   & $\Delta=0.0068(13)$ &$\Delta= 0.0040(7)$& \\
		\hline
        $w_{\rm noz}+w_{\rm sys}$  & $\Delta= 0.00108(228)$ &$\Delta= 0.00138(156)$& \\
        fibre collisions            & $\Delta=-0.00296(121)$ &$\Delta= -0.00026(92)$& \\
		\hline
	\end{tabular}
\end{table}

\begin{figure}
	\includegraphics[width=\columnwidth]{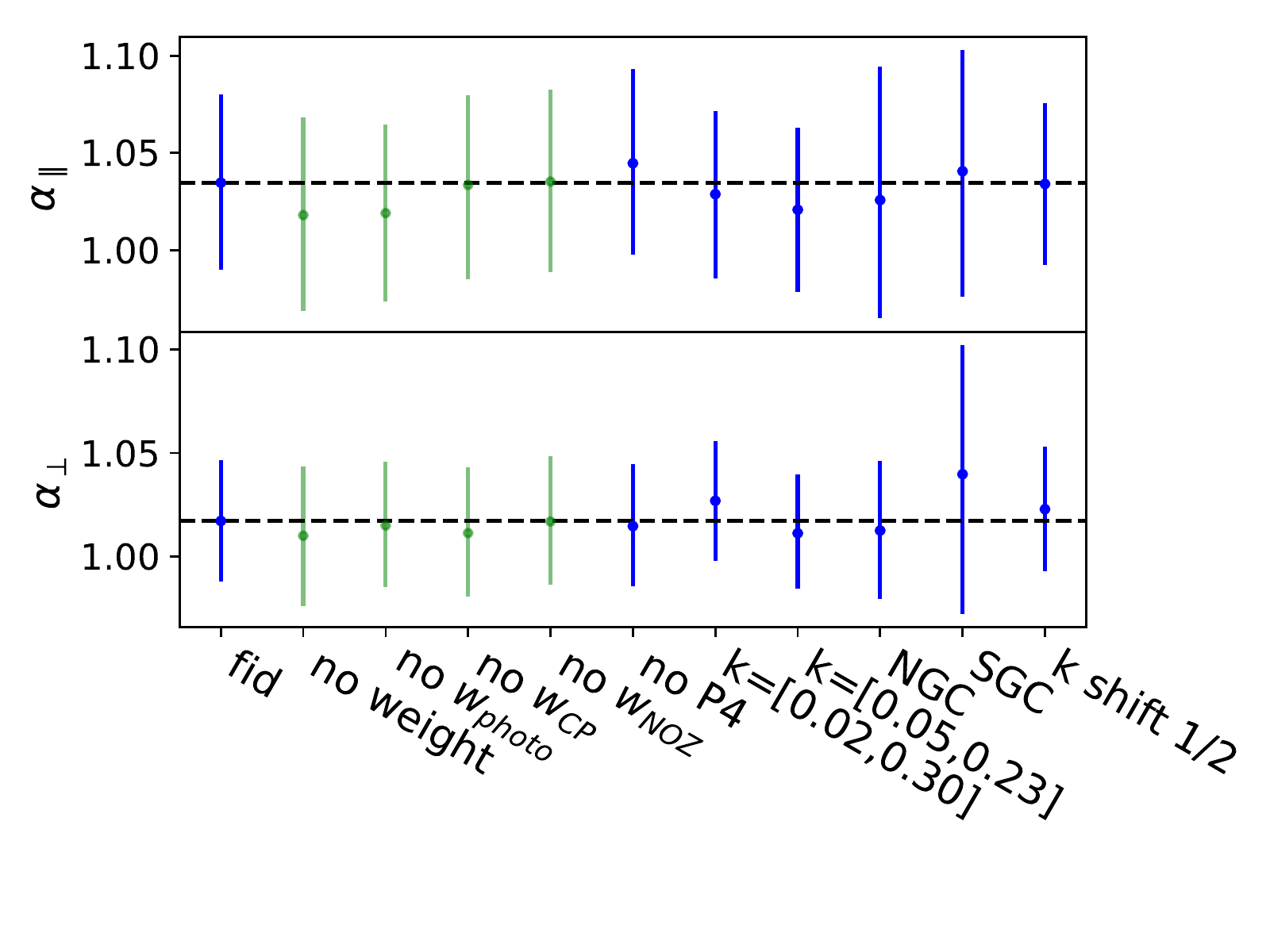}
    \vskip -1.cm
    \caption{Best-fit values of $\alpha_\parallel$ and $\alpha_\perp$ for the tests performed on DR16 quasar sample for the BAO-analysis (values are taken from Table~\ref{results:test-data-bao}). Green points show the impact of taking into account the different weights while blue points are for consistency/robustness tests.}
    \label{fig:test-DR16-BAO}
\end{figure}

The results of systematic checks that were performed on the EZmocks are summarized in Table~\ref{tab:bao-numbers}. 
For these studies, the reference dilation scales are taken from mock catalogues with all observational systematic effect applied and corrected for using the standard weighting scheme (see Eq~\ref{eq:weights-q}). The reference model has the window function (wf) applied and the smooth-term is decoupled from the BAO peak term. Changing the prescription for the window function or smooth-term coupling induces changes which are at maximum 0.7\% (central value averaged over 1000 mocks). The magnitude of the difference is in agreement with the modelling systematic error quoted from the mock challenge (see Eq.~\ref{eq:syst-mc-bao}). 

We quantify the change in dilation scale parameters for different combinations of systematic effects applied to the EZmocks. We show in Sect.~\ref{sec:syst-rsd} 
that the deviations are smooth and can be accounted for by the second degree polynomial broadband terms of the BAO model. As a consequence, the difference between the best fit parameters with and without observational effects is consistent with zero, and we use the statistical error on these offsets (see Table~\ref{tab:bao-numbers}) as a systematic error,
\begin{equation}
\Delta(\alpha_\parallel)\bare{obs}=0.0037 \quad \Delta(\alpha_\perp)\bare{obs}=0.0036.
\label{eq:syst-obs-bao}
\end{equation}

\begin{table*}
	\centering
	\caption{Best fit and 1-sigma error bars for robustness tests performed on the DR16 data for the BAO-only analysis in Fourier space. When the shift with respect to the DR16 final result is larger than 0.01, we indicate in parenthesis () the standard deviation of the mock-by-mock differences.}
	\label{results:test-data-bao}
	\begin{tabular}{lcccccc} 
     &$\alpha_\parallel$ & $\alpha_\perp$ & $\chi^2(n_{\rm dof}$)& red.$\chi^2$\\
\hline
DR16 final result  & $1.035 \pm 0.045$ & $1.017 \pm 0.029$ & 87.63 (126-22) & 0.84 \\
\hline
no wf              & $1.033\pm0.043$ & $1.018\pm0.028$ & (126-22) &  \\ %
coupled            & $1.016\pm0.048$ & $1.016\pm0.031$ & (126-22) & \\ %
                   & (0.029) & (0.017) & & \\
\hline
no weight         & $1.018 \pm 0.050$ & $1.010 \pm 0.034$ & 70.86 (126-22) & 0.68 \\
no $w_{\rm sys}$  & $1.019 \pm 0.045$ & $1.015 \pm 0.030$ & 85.32 (126-22) & 0.82 \\
no $w_{\rm cp}$   & $1.034 \pm 0.047$ & $1.011 \pm 0.031$ & 78.36 (126-22) & 0.75 \\
no $w_{\rm noz}$  & $1.035 \pm 0.047$ & $1.017 \pm 0.031$ & 79.90 (126-22) & 0.77 \\
\hline
no $\ell=4$          & $1.045 \pm 0.048$ & $1.015 \pm 0.030$ & 46.39 (84-16) & 0.68 \\
                  & (0.017) & (0.013) & & \\
\hline
k=[0.02,0.30]  & $1.029 \pm 0.043$ & $1.027 \pm 0.029$ & 107.46 (168-22) & 0.74 \\
               & (0.017) & (0.014) & & \\
k=[0.05,0.23]  & $1.021 \pm 0.042$ & $1.011 \pm 0.028$ & 71.57 (108-22) & 0.83 \\
 & (0.036) & (0.022) & & \\
k shift 1/2  & $1.034 \pm 0.042$ & $1.023 \pm 0.030$ & 92.41 (126-22) & 0.89 \\
\hline
$\Sigma_s=4-2\,{\rm Mpc/h}$ & $1.031\pm0.042$ & $1.016\pm0.029$ & 86.06 (126-22) &\\
$\Sigma_s=4+2\,{\rm Mpc/h}$ & $1.040\pm0.049$ & $1.017\pm0.030$ & 89.50(126-22) & \\
\hline
NGC  & $1.026 \pm 0.065$ & $1.013 \pm 0.033$ & 46.63 (63-12) & 0.91 \\
SGC  & $1.041 \pm 0.063$ & $1.040 \pm 0.065$ & 41.12 (63-12) & 0.81 \\
\hline
isotropic BAO & \multicolumn{2}{c|}{$\alpha_\mathrm{iso} = 1.025 \pm 0.020$} & 26.76 (42-9) & 0.81 \\
\hline
\end{tabular}
\end{table*}


We perform further robustness tests on the data and the results are summarized in Table~\ref{results:test-data-bao} and displayed in Fig.~\ref{fig:test-DR16-BAO}. When the observed shift in dilation scale parameters are larger than 1\%, we compare them with the standard deviation of the mock-by-mock differences (see values in parenthesis in Table~\ref{results:test-data-bao}). 


We measure the difference in the best fit parameters between the decoupled and the coupled smooth-term prescriptions, and we observe variations of the order of $0.019$ for $\alpha_\parallel$ and $0.001$ for $\alpha_\perp$. Using the EZmocks, we found that the standard deviations of the mock-by-mock differences are $\sigma_{\rm mocks}(\alpha_\parallel)=0.029$ and $\sigma_{\rm mocks}(\alpha_\parallel)=0.017$. Therefore, the observed variation is within statistics and we do not assign any additional systematic error to cope with this effect.

Then, we show the impact of the different weights on the cosmological parameters estimation. It appears that taking into account photometric weight has the largest impact although the overall effect is smaller than half the statistical precision. The fibre collision and spectroscopic redshift weights have only a marginal effect on the best fit parameters. 
Not taking into account the hexadecapole in the fit changes $\alpha_\parallel$ by $0.010$ and $\alpha_\perp$ by $0.02$. 

Changing the fitting range for the BAO analysis is also studied. First, the upper bound is increased to $k=0.3\hMpc{}$ bringing in scales for which BAO oscillations are no longer visible in the data. Adding these data produce an effect of $\Delta \alpha_\perp = 0.010$ which is the largest deviation in the tests that were done for this parameter. It is due to the fact that the added data put a stronger constraint on the broadband terms in a region without BAO signal, and therefore removes the ability of the model to account for broad band variations in the region of higher BAO significance. Adding more terms in the broadband polynomial expansion could relieve this effect but this goes beyond the validation of the model that was performed in the mock challenge.
Removing scales below $k<0.05\hMpc{}$ has a noticeable effect on the radial dilation scale $\Delta \alpha_\parallel = 0.014$ as it removes scales where the amplitude of the BAO wiggles is large as can be seen for the mocks in Fig.~\ref{fig:bao_wiggle}. The differences when changing the upper or lower bound of the $k$-range are within one standard deviation of the differences observed in the EZmocks. Shifting the $k$-bins by half the bin width ($\Delta k=0.005\hMpc{}$) has a minor impact. 

We also study the variation in the best fit parameters when changing the strength of the damping terms that are determined using the OuterRim mocks. We observe variations that are at the level of 0.1\% for extreme changes of $\pm30$\% in $\Sigma_\parallel$ and $\Sigma_\perp$. In the case of $\Sigma_s$, the sensitivity on $\alpha_{\parallel}$ is larger, reaching $0.5$\% for variations of $\Delta \Sigma_s =\pm50$\% as shown in the bottom part of Table~\ref{results:test-data-bao}. The observed changes in cosmological parameters are taken as a systematic error due to the knowledge of the damping strength:
\begin{equation}
\Delta(\alpha_\parallel)\bare{damping}=0.005\quad \Delta(\alpha_\perp)\bare{damping}=0.001.
\label{eq:syst-damping-bao}
\end{equation}
Furthermore, the error on $\alpha_{\parallel}$ is affected by the change in the damping term as our data have a sharp BAO feature and the fit prefers less damping. The statistical errors that we measure using our fiducial choice of damping parameters are close to the average of the errors on $\alpha_\parallel$ and $\alpha_\perp$ for the various cases that were studied.

The fit was also performed for each galactic cap separately and the differences observed are within the variations expected from the statistics. One should note that, because the strength of the BAO for each cap is different, the precision on the best fit parameters does not follow the difference in surface area of each sub-sample (see Table~\ref{tab:area}). 

The systematic errors on $\alpha_\parallel$ and $\alpha_\perp$ for the BAO analysis are summarized in Table~\ref{tab:sys-tot-bao}. The dominant contribution stems from the error in the modelling. Adding systematic errors contributions in quadrature, we obtain a 1.2\% error on $\alpha_\parallel$ and 0.7\% on $\alpha_\perp$. These errors represent approximately 25\% of the statistical errors.

\begin{table}
	\centering
	\caption{Systematic errors on the estimate of the cosmological parameters from the BAO analysis.}
	\label{tab:sys-tot-bao}
	\begin{tabular}{lccc} 
		\hline
		   & $\alpha_\parallel$ & $\alpha_\perp$\\
		\hline
		observational & 0.0037 & 0.0036 \\
		modelling     & 0.0098 & 0.0055 \\
		damping       & 0.005  & 0.001 \\
		\hline
		total systematics    & 0.012 & 0.007 \\
		statistical error    & 0.045 & 0.029 \\
		fraction             & 27\%  & 24\% \\
		\hline
\end{tabular}
\end{table}

\subsubsection{Results from the BAO analysis and consensus}

\begin{figure}
	\includegraphics[width=\columnwidth]{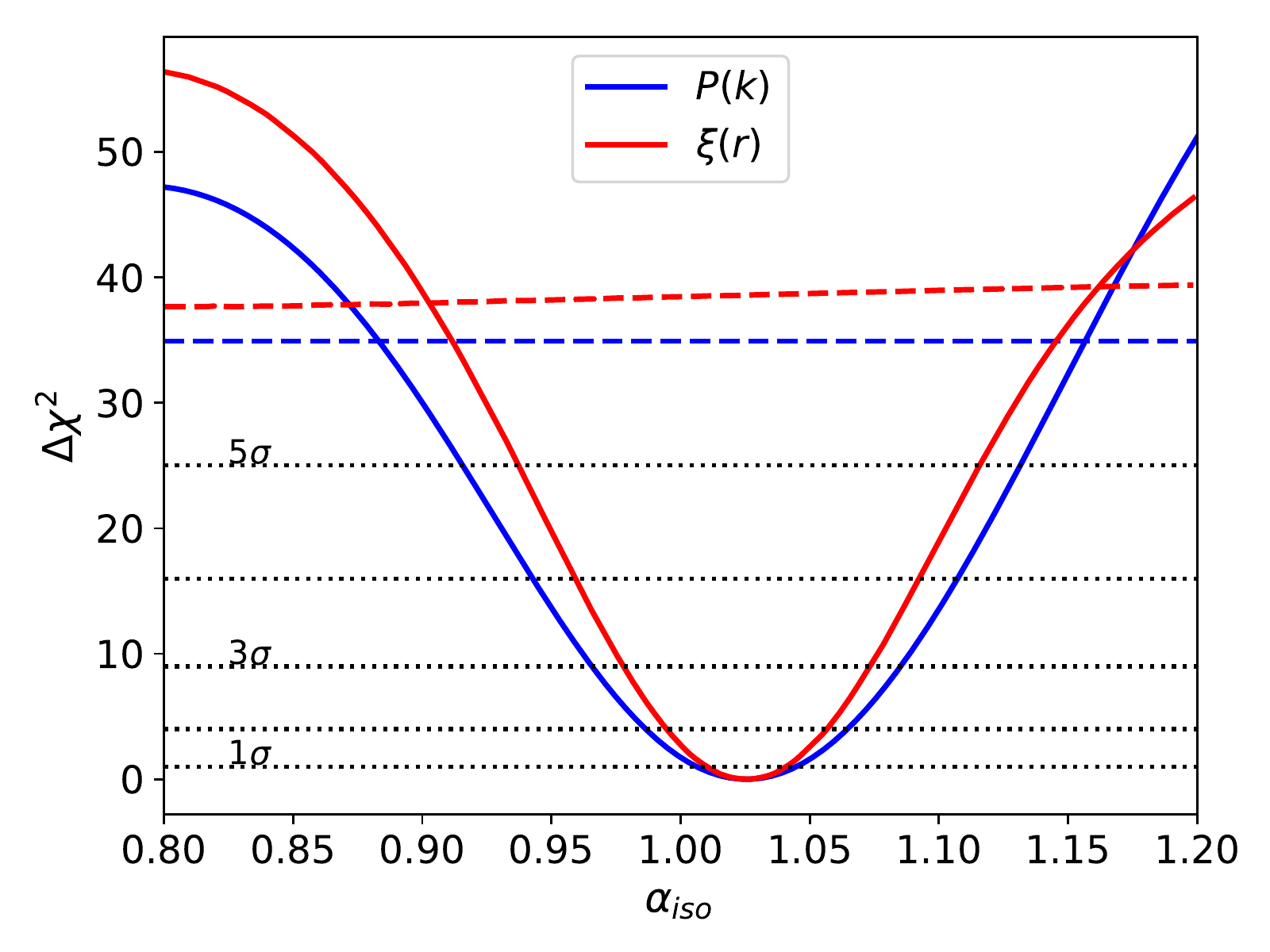}
    \caption{$\chi^2$ profile of the $\alpha_{\rm iso}$ BAO parameter in Fourier and configuration space. We show the $\chi^2$ profile for the BAO model (solid curves) and the $\chi^2$ difference between a model without BAO peak and the minimum of the BAO model (dashed lines)}
    \label{fig:chi2-DR16-BAO}
\end{figure}

The results of our analysis are compared to the BAO analysis performed in configuration space, which is described in our companion paper~\citep{Hou+20}. In Fig.~\ref{fig:chi2-DR16-BAO}, we show the variation of the minimum $\chi^2$ of our model as a function of the assumed isotropic dilation scale $\alpha_\mathrm{iso}$ and compare it with the $\chi^2$ for the model without BAO oscillations. This shows that our data confirm the presence of the BAO signal at the 5- to 6-$\sigma$ level, in agreement with the results obtained in configuration space~\citep{Hou+20}.

\begin{figure*}
	\includegraphics[width=1.2\columnwidth]{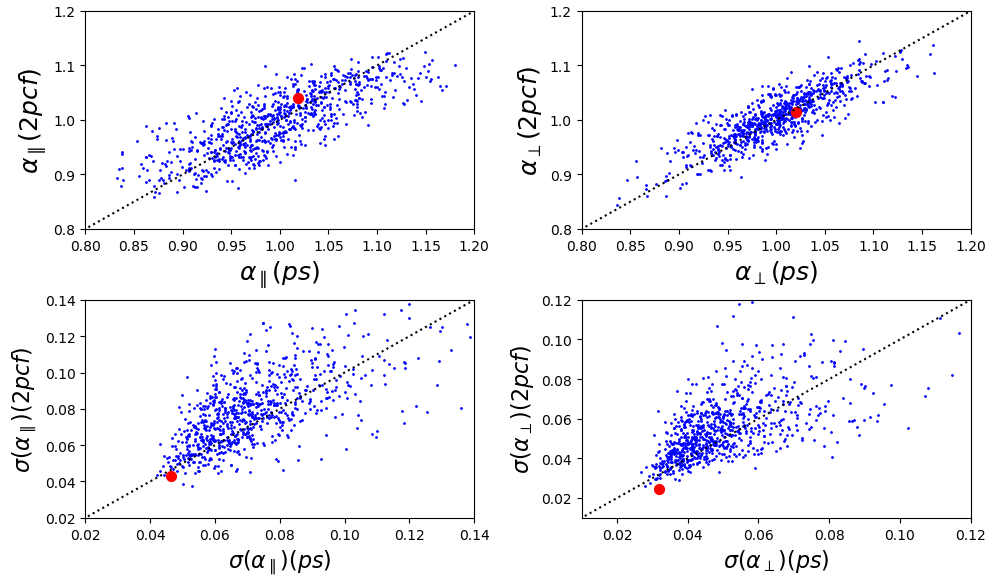}
	\includegraphics[width=0.66\columnwidth]{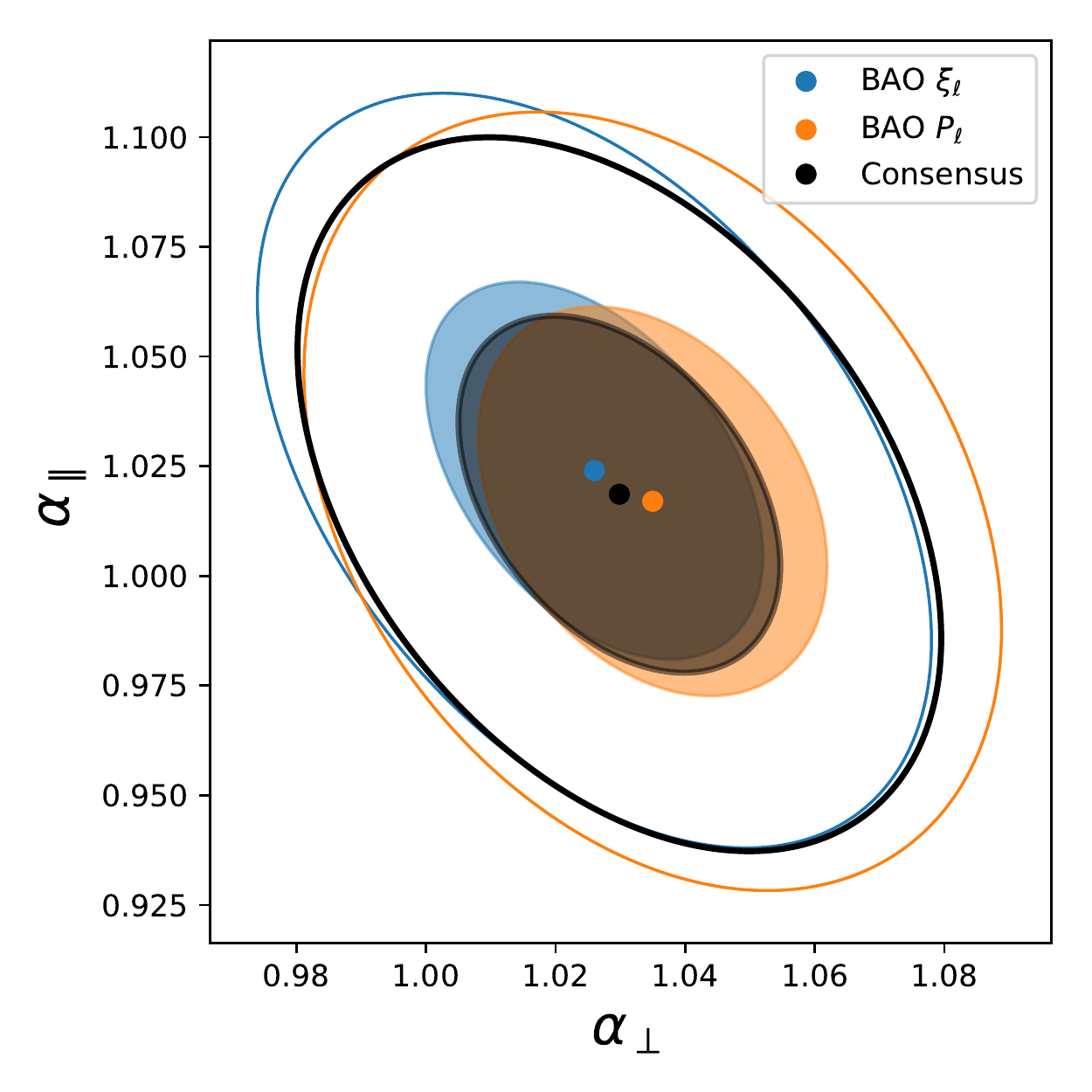}
    \caption{Left: Comparison of the cosmological parameters and errors measured in the BAO analysis of the 2-point correlation function analysis, from~\citet{Hou+20}, (vertical axis) and of the power spectrum analysis (horizontal axis). The blue points show values of the EZmocks fit while the red point stands for the DR16 measurement. Right: Likelihood contour for the 2 analyses and the consensus. The orange (blue) contours represent the power spectrum (2-point correlation function) analysis; the black contours represent the consensus}
    \label{fig:bao-compare-scatter}
\end{figure*}

In Fig.~\ref{fig:bao-compare-scatter} we compare the parameters measured in configuration and Fourier space for the 1000 approximate mocks. After selecting mocks for which there is a clear detection of the BAO signal in either analysis ( the selection criteria is $0.82<[\alpha_\parallel,\alpha_\perp]<1.18$ keeping 742/1000 mocks) the Pearson correlation coefficients reach $\rho(\alpha_\parallel)=$0.795 and $\rho(\alpha_\perp)=0.821$. The errors in $\alpha_\parallel$ and $\alpha_\perp$ obtained in configuration space are comparable to the errors from the power spectrum fits, although the errors in configuration space on average tend be larger in the low S/N regime. The DR16 measurements are shown by the red points in the left panel Fig.~\ref{fig:bao-compare-scatter}. The errors measured from the DR16 data are at the edge of the distribution of the EZmocks for both analyses. This is expected since the BAO signal observed in the data is stronger than the average BAO signal in the mocks, as was shown in Fig.~\ref{fig:bao_wiggle}.

In the right panel of Fig.~\ref{fig:bao-compare-scatter}, we compare the likelihood contours obtained for the Fourier and configuration space BAO analyses as well as the consensus result. The results are in good agreement, as the difference between the two represents only 30\% of the standard deviation of the mock-by-mock differences for both $a_\parallel$ and $a_\perp$.

Then, following~\citet{Sanchez+2017}, we perform a consensus analysis by computing :
\begin{equation}
    D_c=\Psi_c^{-1}\sum_{i=1}^m \sum_{j=1}^m\Psi_{tot,ji}D_i,
\end{equation}
where $D_c$ is the parameter consensus vector, $\Psi_c$ is the inverse of the consensus covariance matrix, $m$ is the number of different statistical analyses (in this case $m=2$) and $\Psi_{tot}$ is the inverse of the total covariance matrix (each $\Psi_{tot,ji}$ term is a $p\times p$ matrix with $p$ the number of parameters). The cross covariance matrix elements are measured using the approximate mocks while the diagonal comes from the the covariance matrix of the individual data fits. We convert the dilation scale parameters to the cosmological parameters $D_{\rm H}/r_{\rm drag}$ and $D_{\rm M}/r_{\rm drag}$ with eq.~\ref{eq:cosmo_dilationscale}
The final results of these cosmological parameters for the BAO analysis in Fourier and configuration spaces and their consensus are given in Table~\ref{tab:consensusBAO}. 
We present the consensus covariance matrix including the statistical errors, the modelling systematics, and the observational systematics; 
\begin{equation}
\mathbf{C}_{\rm bao} = \begin{blockarray}{cc}
D_{\rm H}/r_{\rm drag} & D_{\rm M}/r_{\rm drag} \smallskip\\
\begin{block}{( cc )}
  0.3047 & 0.1707 \\ 
  - & 0.6373 \\
\end{block}
\end{blockarray}
\end{equation}
The consensus results are used to derive the cosmological implications of the eBOSS data~\citep{Eva}.

\begin{table*}
\caption{Final results of the BAO-only analyses in Fourier and configuration spaces and their consensus.}
\begin{center}
\begin{tabular}{|c|c|c|c}
 & $D_{\rm H}(z_{\rm eff})/r_{\rm drag}$ & $D_{\rm M}(z_{\rm eff})/r_{\rm drag}$ & $D_V(z_{\rm eff})/r_{\rm drag}$\\
\hline
\hline
Fourier Space & $13.34 \pm 0.60$ & $30.60 \pm 0.90$ & $26.50 \pm 0.55$\\
Configuration Space & $13.22 \pm 0.58$ & $30.82 \pm 0.85$ & $26.52 \pm 0.44$\\
\hline
{\bf BAO-only Consensus} & \boldmath$13.26 \pm 0.55$ & \boldmath$30.69 \pm 0.80$ & \boldmath$26.51 \pm 0.42$\\
\hline
\end{tabular}
\end{center}
\label{tab:consensusBAO}
\end{table*}

\section{Full Shape RSD analysis}
\label{sec:FS}

In this section, we present the Full Shape RSD analysis of the eBOSS DR16 quasar power spectrum. First, we briefly describe the power spectrum model, then we present the various tests performed both on the EZmocks and on the data to estimate the systematic errors in our measurement. Finally, we present the results we obtain and perform a consensus analysis with the measurement in the configuration space for the same sample as presented in~\citet{Hou+20}.

\subsection{Model}
\label{sec:model}
The RSD model used in this analysis is based on the work of~\citet{TNS10}. We use the implementation of~\citet{DeMattia+20} and we refer the reader to section~3.1 of this paper for further details. In this model, the non linear power spectrum for a tracer of the matter distribution, with bias $b_1$, is given by:
\begin{equation}
\begin{split}
P_q(k,\mu) =& P_{q,\delta\delta}(k)+2f\mu^2P_{q,\delta\theta}(k)+f^2\mu^4P_{q,\theta\theta}(k)\\
           &+ b_1^3A(k,\mu,f/b_1)+b_1^4B(k,\mu,f/b_1)
	\label{eq:tns}
\end{split}
\end{equation}
where $P_{q,\delta\delta}$ $P_{q,\delta\theta}$ and $P_{q,\theta\theta}$ are the quasar-quasar, quasar-velocity and velocity-velocity power spectra respectively, and $A$ and $B$ are correction terms to account for the non linearity of the real to redshift space mapping. The 1-loop bias terms are taken from~\citet{McDonald+2009} and all terms are calculated at 2-loop order following the \texttt{RegPT} scheme~\citep{regpt}.It is worth noting that in the bias expansion of $P_{q,\delta\delta}$ there is an additional term, $N_g = A_g P_0^{\rm noise}$, to account for the constant galaxy stochasticity.

The power spectrum of the quasars, ${\cal P}_q(k,\mu)$, is the product of the non linear power spectrum $P_q(k,\mu)$ (defined in Eq.~\ref{eq:tns}) for a biased tracer and of a damping term $D$ which takes into account non linear effects that are not included in the matter power spectrum, 
\begin{equation}
    {\cal P}_q(k,\mu)=P_q(k,\mu)D(k,\mu).
	\label{eq:model}
\end{equation}
Following~\citet{Hou+18}, we use a damping term which comprises a Gaussian and a Lorentzian-like term $a_{\rm vir}$ ,
\begin{equation}
    D=\frac{1}{\sqrt{1+(k\mu a_{\rm vir})^2}}\exp{\left[-\frac{(k\mu \sigma_v)^2}{1+(k\mu a_{\rm vir})^2}\right]},
	\label{eq:fog}
\end{equation}
where $\sigma_v$ is the velocity dispersion, and $a_{\rm vir}$ is associated with the virial motion of quasars in the dark matter halo they inhabit. In the approach adopted here, we consider redshift errors as an additional velocity dispersion, and we let $\sigma_v$ vary as a free parameter in order to cope for this effect. We do not add an extra term as in~\citet{Hou+18}.

Furthermore, the window function measured from the data is applied to the model following the same method that was used for the BAO analysis.

For the fiducial analysis, the fit is performed over the range $k=[0.02, 0.3]\hMpc$ and 13 parameters are allowed to vary. The cosmological parameters ($\alpha_{\perp}$, $\alpha_{\parallel}$, $f$) are common to the 2 galactic caps, while the parameters of the bias expansion ($b_1$,$b_2$), the shotnoise term ($A_g$) and of the damping term ($\sigma_v$,$a_{vir}$) are allowed to be different for the 2 galactic caps. We use flat priors for all parameters, and the intervals of variations are given in Table~\ref{tab:fs-prior} and are chosen such that the boundaries are not hit.

\begin{table}
	\centering
	\caption{Interval of variations of the parameters used in the $\chi^2$ minimisation for the Full Shape RSD analysis.}
	\label{tab:fs-prior}
	\begin{tabular}{ll} 
		\hline
		       parameter   & prior range \\
		\hline
		$b_1$          &  $[0,   5] $\\
		$b_2$          &  $[-8,  8] $\\
        $a_\parallel$  &  $[0.5, 1.5]$ \\
		$a_\perp$      &  $[0.5, 1.5]$ \\
		$f$            &  $[0.3, 3] $\\
		$A_g$          &  $[-1,  5] $\\
		$\sigma_v$     &  $[0,   15]$ \\
		$a_{vir}$      &  $[0,   15]$ \\
		\hline
	\end{tabular}
\end{table}

\subsection{Validation of the model}

We validate our model using the OuterRim mocks described in Section~\ref{sec:ORmock}. This is described in detail in the companion paper of~\citet{Smith+20} and we only give the main results here. In a first stage, we used the non-blind mocks, which include various redshift smearing prescriptions to test the damping term, $D$, introduced in the previous paragraph. We fit our power spectrum model to the 100 realisations of each HOD model, and compute the average of the best fit parameters. The results show that the true values of $\alpha_\parallel$ and $\alpha_\perp$ can be recovered to better than 1\% and that $f\sigma_8$ can be recovered to better than 3\% regardless of the redshift smearing prescription. 

In a second stage, we analysed the `blind' set of mocks, whose cosmology is unknown, using the OuterRim fiducial cosmology. The results of the average fit parameters for the 24 sets of mocks with different HOD and different cosmologies are given in section~6.3 of~\cite{Smith+20}. From the distribution of the results that we obtain, we determine the rms value to be our systematic error on the modelling:
\begin{equation}
\Delta \alpha_\parallel\bare{model}=0.0098 \quad  \Delta \alpha_\perp\bare{model}=0.0066 \quad\Delta f\sigma_8\bare{model}=0.0123
	\label{eq:fs-model-syst-errors}
\end{equation}

\subsection{Results}

\begin{figure*}
	\includegraphics[width=\columnwidth]{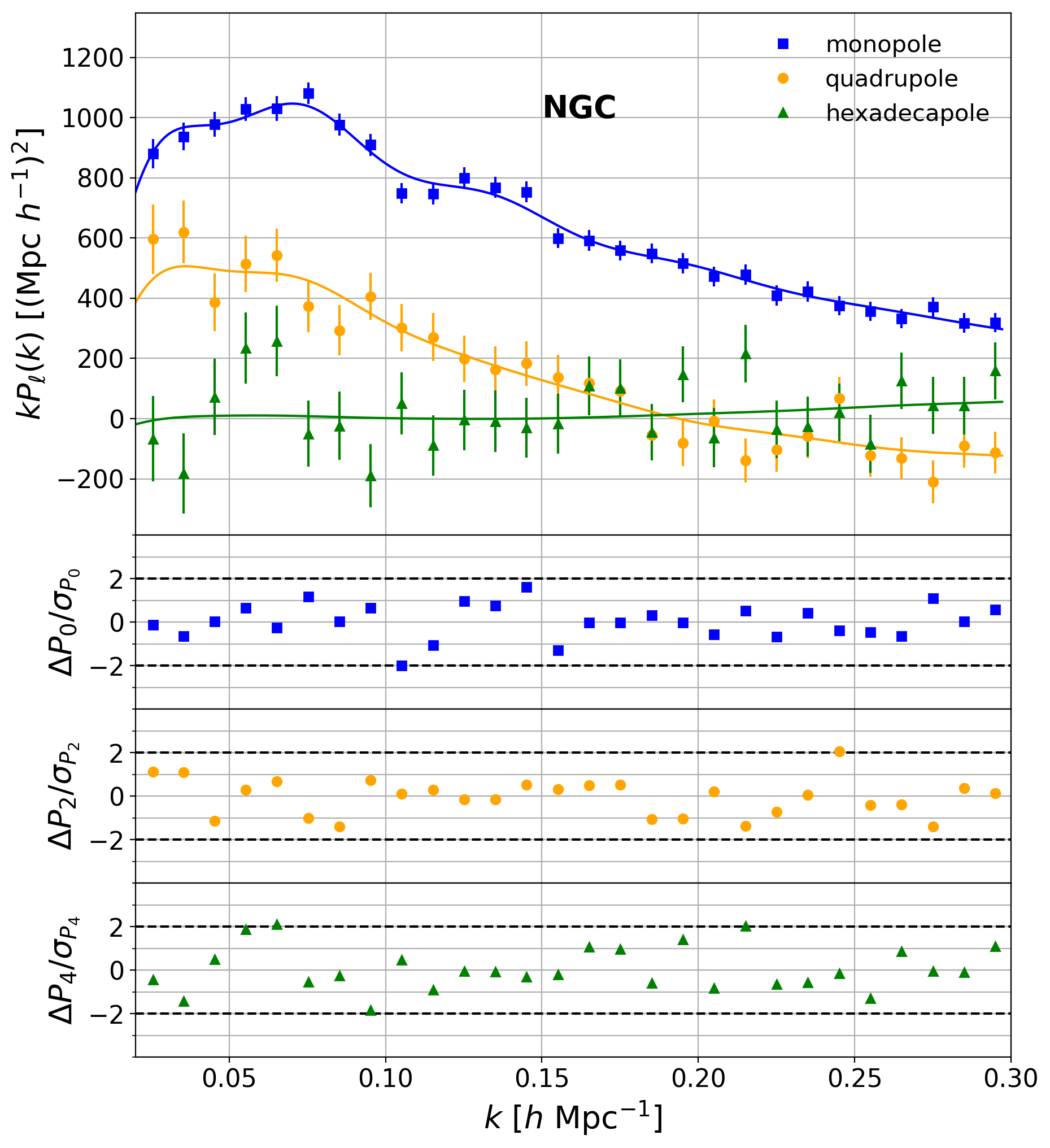}
	\includegraphics[width=\columnwidth]{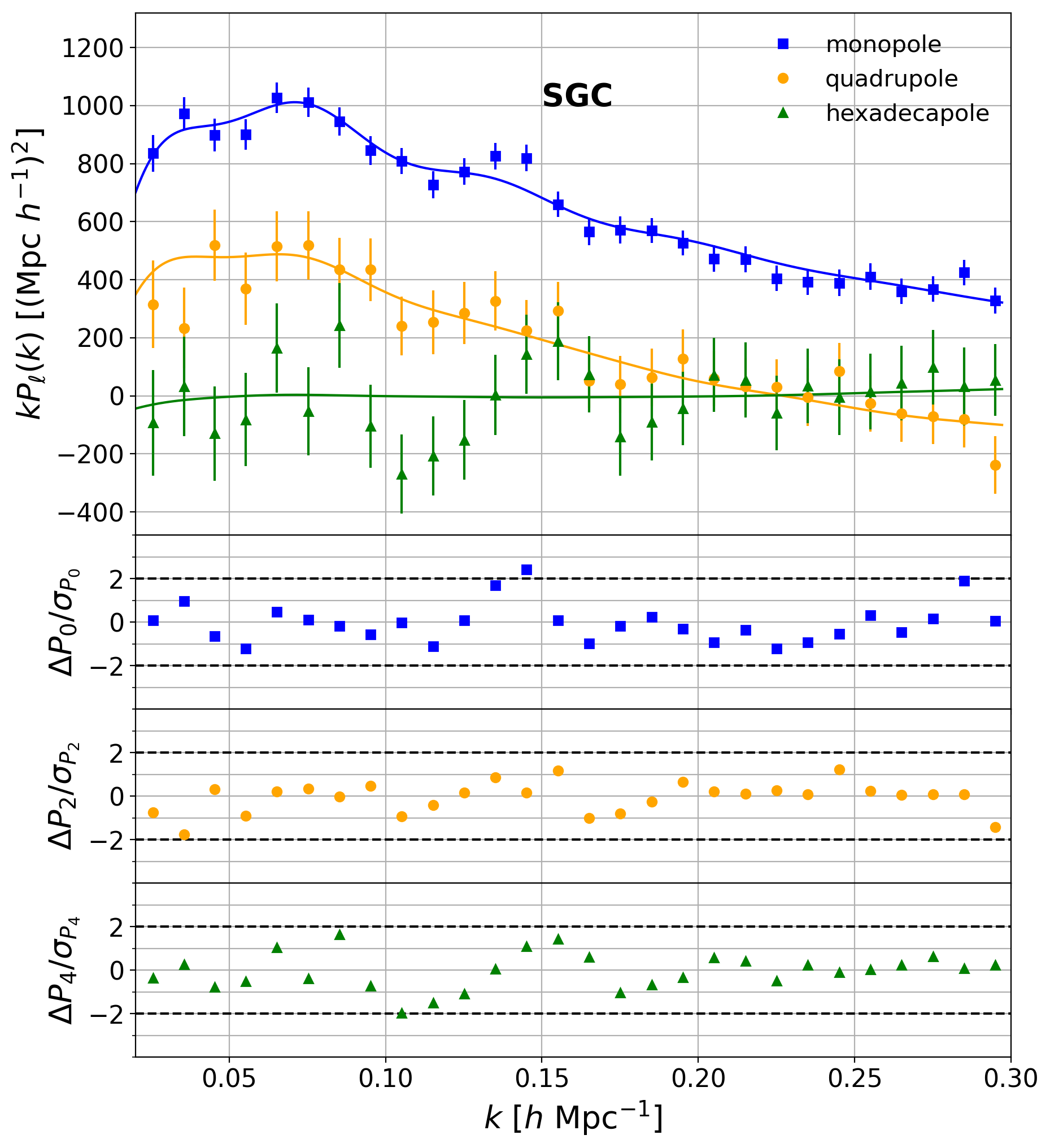}
    \caption{Power spectrum multipoles of the NGC (left) and SGC (right) quasar samples (top panel) and residuals (lower panels) from the NGC and SGC combined fit for the Full-Shape RSD analysis. The points are the data, and the solid lines show the best fit model.}
    \label{fig:ps-ngc-sgc}
\end{figure*}

We perform the Full Shape RSD analysis of the monopole, quadrupole, and hexadecapole of the power spectrum measured from the QSO DR16 catalogue, as described in Section~\ref{sec:model}. The covariance matrix is computed from the 1000 EZmocks. Fig.~\ref{fig:ps-ngc-sgc} shows the best fit model (solid lines) to the data (points) with NGC and SGC fitted simultaneously. We take the mean of the samples of the Markov chains as final results for the full shape RSD analysis in Fourier Space:
\begin{equation}
\alpha_\parallel=1.049 \pm 0.038 \quad \alpha_\perp=1.020 \pm 0.029 \quad f\sigma_8=0.476 \pm 0.044,
\label{eq:fs-results}
\end{equation}
where the errors are computed from the standard deviation of the samples of the chains. In this section, we discuss the evaluation of the systematic errors entering our measurement. We use the EZmocks to quantify the impact of the different weights (fibre collisions, photometric conditions, redshift efficiency). Fibre collisions weights have a large impact and require an improvement of the power spectrum model that is presented. Taking into account the radial integral constraint induced by the generation of the random catalogue is also evaluated.
At the end of the section, we present the results of robustness tests performed on the DR16 data and show the impact of the different analysis choices on the final results.
For the estimation of the systematic errors using the 1000 approximate mocks, we measure the distributions of the mock by mock differences. These distributions are fit with a Gaussian from which the mean values and their errors are determined assuming the differences distributions follow Poisson statistics. We take the largest quantity between the mean value or twice the error as a systematic error. 

\begin{table}
	\centering
	\caption{Average value of the cosmological parameters recovered from the fits of 1000 EZmocks under different systematic effects applied to the catalogues and corrected for using the standard weighting scheme. The lines that start with a $\Delta$ show the difference in offset between the preceding line with respect to the offset measured for the line described in parenthesis. The "expected" values for the dilation scales are not unity since the EZmocks cosmology is slightly different from the fiducial cosmology.}
	\label{tab:sys-fs}
	\begin{tabular}{lccc} 
		\hline
		 tests on mocks   & $\alpha_\parallel$ & $\alpha_\perp$ &$f\sigma_8$\\
		\hline
		expected     & {\it 1.004} & {\it 1.002} & {\it 0.379} \\
           ref : no weights & 0.988$\pm$0.002 & 0.990$\pm$0.001 &  0.380$\pm$0.002 \\
$w_{\rm noz}w_{\rm sys}$ & 0.988$\pm$0.002 & 0.992$\pm$0.001 &  0.382$\pm$0.002 \\
              $\quad \Delta_1 $ wrt (ref) & -0.0006(15) & 0.0019(12) & 0.0026(14) \\
$w_q=w_{\rm noz}w_{\rm sys}w_{\rm cp}$     & 0.974$\pm$0.002 & 0.999$\pm$0.001 &  0.399$\pm$0.002 \\
 $\, \Delta $ wrt ($w_{\rm noz}w_{\rm sys}$) & -0.0147(9) & 0.0081(7) & 0.0166(9) \\ 
        \hline
	$w_q+\Delta P_{cp}^{u}$   & 0.991$\pm$0.002 & 0.989$\pm$0.001 &  0.380$\pm$0.002 \\
 $\, \Delta_2 $ wrt ($w_q$)	& 0.0172(6) & -0.0107(6) & -0.0205(6) \\
        \hline
	$w_q+\Delta P_{\rm cp}^{u}+\Delta P_{\rm cp}^{c}$   & 0.991$\pm$0.002 & 0.990$\pm$0.001 &  0.381$\pm$0.002 \\
 $\, \Delta $ wrt ($w_q+\Delta P_{\rm cp}^{u}$)  & 0.0008(6) & 0.0007(5) & 0.0024(6) \\
 $\, \Delta_3 $ wrt ($w_{\rm noz}w_{\rm sys}$)        & 0.0040(9) & -0.0018(7) & -0.0011(10) \\
 $\, \Delta $ wrt (ref)         & 0.0056(15) & -0.0000(13) & 0.0007(15) \\
        \hline
no RIC                    & 0.995$\pm$0.002 & 0.986$\pm$0.001 &  0.381$\pm$0.002 \\
$\, \Delta_4 $ wrt (ref)  & 0.0074(8) & -0.0039(6) & 0.0004(8) \\
        \hline
RIC corrected             & 1.001$\pm$0.002 & 0.985$\pm$0.001 &  0.383$\pm$0.002 \\
$\, \Delta_5 $ wrt ($w_q+\Delta P_{\rm cp}^{u}$)  & 0.0077(1) & -0.0047(1) & 0.0015(1) \\
\hline
	\end{tabular}
\end{table}

\subsubsection{Systematic checks}
\label{sec:syst-rsd}

\begin{figure*}
	\includegraphics[width=1.\columnwidth]{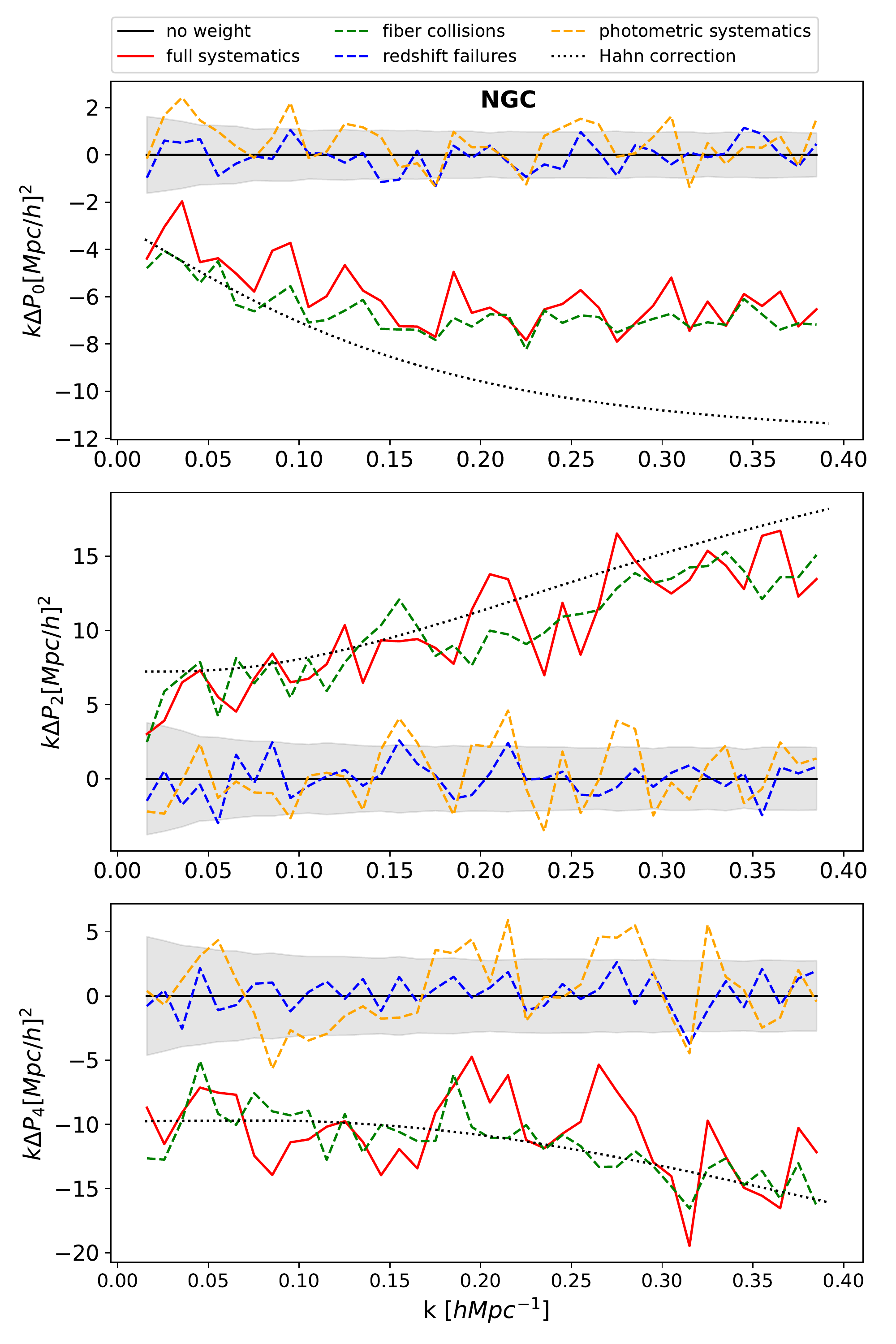}
	\includegraphics[width=1.\columnwidth]{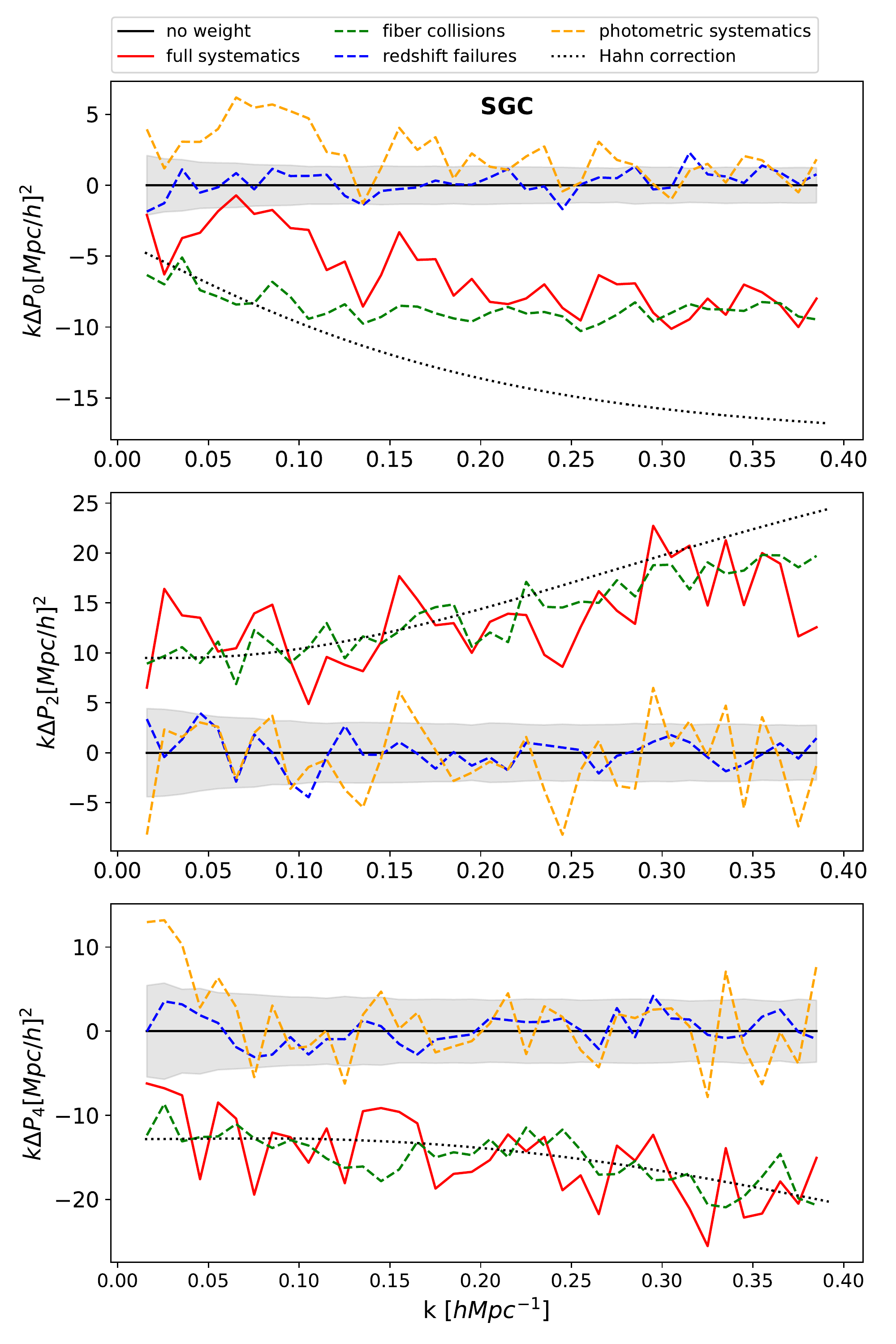}
    \caption{Shifts on the power spectrum multipoles induced by the systematic effects applied to the EZ mocks on the NGC (left panels) and SGC (right panels). The dashed line represents the effect of fibre collisions weights (green), of photometric systematic weights (yellow) and redshift failures weights (blue). The effect of all weights together is shown as the red solid line. The black dotted lines show the full correction for fibre collision as proposed by~\citet{Hahn+2017}. The grey shaded region represent the statistical error.}
    \label{fig:ps-hahn}
\end{figure*}

In Fig.~\ref{fig:ps-hahn}, we show the change in the the power spectrum multipoles when different combinations of systematic effects are included. In each case, the effects are applied to the EZmock catalogues and are corrected for according to the weighting scheme used on the data. It appears that largest systematic offset arises from fibre collisions, and no difference between the two galactic caps is observed beyond the expected statistical error. The impact of the systematic effects on the best fit parameters is shown in Table~\ref{tab:sys-fs} for the average of the 1000 EZmocks and for the combined NGC+SGC fit. In this table, the lines that start with a $\Delta$ show the difference in offset between the preceding line with respect to the offset measured for the line described in parenthesis. It shows that not correcting for fibre collisions leads to large systematic offsets on all cosmological parameters that go up to 5\% for the case of the $f\sigma_8$.

There is an effect of imperfectly correcting for photometric conditions which affects the monopole (yellow dashed curves in Fig.~\ref{fig:ps-hahn}). The shift is located at small $k$ and amounts to about 1$\%$ of the observed monopole and no effect beyond statistics is observed in the higher order multipoles. The shift is higher in the SGC for which the spread of the photometric weights is known to be larger than the NGC (see figure~12 of~\citealt{Zarrouk+18}). The best fit parameters are slightly modified by the amount given in the line $\Delta_1$ of Table~\ref{tab:sys-fs} 
that shows the impact of photometric weights and redshift failures together. This difference
is then taken as an estimate of the systematic errors arising from photometric conditions. 

Taking into account fibre collisions has received much attention, and solutions have been proposed to mitigate their effect on the power spectrum~\citep{Hand+2017} and on the 2-point correlation function~\citep{PIPPercivalB2017}. In the approximate method proposed by~\citet{Hahn+2017} that we use in this paper, the fact that two targets are colliding is modelled by an additional top-hat window function whose width is given by the collision radius $D_{\rm fc}$ at the effective redshift ($D_{\rm fc}(z=1.480)=0.9\Mpch$). In the case the fibre collided objects are uncorrelated, the corrected power spectrum is given by:
\begin{equation}
    \Delta P_\ell^{\rm uncorr}(k)=-f_s (2l+1)L_\ell(0) \frac{(\pi D_{\rm fc})^2}{k} W_{2D}(kD_{\rm fc}),
	\label{eq:hahnuncorr}
\end{equation}
where $W_{x}=2J_1(x)/x$ is the cylindrical top-hat function in 2D ($J_1$ is the first kind and first order Bessel function), and $f_s$ is the fraction of the survey affected by fibre collisions. The latter is determined from the data targets and observational catalogues. Its estimation requires running the tiling algorithm on the full target set, which includes `legacy' objects that have been observed prior to eBOSS, and comparing this to the number of collisions that were actually corrected for in our sample. Our estimation leads to
\begin{equation}
    f_s(NGC)=0.36 \quad  f_s(SGC)=0.45.
	\label{eq:fs}
\end{equation}
It is substantially lower than the fraction of area of the survey where plates are non-overlapping and collisions are never resolved (60\%). It originates from the fact that the `legacy' objects were themselves the result of a tiling algorithm, where collisions were partially resolved. This estimate of $f_s$ was later confirmed to a precision of ~10\% from the data by measuring the projected correlation function at small transverse scales (see Fig. 16 of~\citealt{Hou+20}).

Using the EZmocks, we estimate that for the target density of our sample, $95\%$ of collisions are uncorrelated. Given that the small-scale clustering of the EZmocks is imperfect, we have confirmed this estimate with the data as it constitutes the dominant part of the correction applied. Nevertheless, as proposed by~\citet{Hahn+2017}, we apply the correction for collisions in the case of correlated objects as:
\begin{equation}
    \Delta P_\ell^{\rm corr}(k)=-f_s \pi D_{\rm fc}^2 \int \frac{d^2q_{\perp}}{(2 \pi)^2} P(k_{\parallel},q_{\perp})W_{2D}(q_{\perp}D_{\rm fc}).
	\label{eq:hahncorr}
\end{equation}
This correction further decreases the residual offset seen in $f\sigma_8$ and does not modify the offset on $a_\parallel$ and $a_\perp$ (Table~\ref{tab:sys-fs}). The remaining offset after both corrections are applied (line $\Delta_3$ of Table~\ref{tab:sys-fs}) is taken as the systematic error due to fibre collisions. In a companion paper,~\citet{Mohammad+20} explores the possibility to use the pair weighting technique~\citep{PIPPercivalB2017} for our sample. This method was not used in the present work but its potential for future surveys is indubitable.

The size of the correction given in Eq.~\ref{eq:hahncorr} is shown as a dotted line in Fig.~\ref{fig:ps-hahn} that is qualitatively in agreement with the observed systematic shift for all multipoles, and also captures the difference between the NGC and the SGC. The agreement is a little worse for the monopole, but this is negligible, since the shift in the monopole is very small compared to its amplitude. After applying this correction to the power spectrum model, the systematic offsets in the best fit parameters measured from the EZmocks are reduced by a factor of ~5 (Table~\ref{tab:sys-fs}), to an acceptable level of the order of one tenth of the statistical error on each parameter. The correction depends linearly on the value of $f_s$ that is known to a precision of $10\%$. Therefore, we take 10\% of the shifts due to this effect (line $\Delta_2$ of Table~\ref{tab:sys-fs}) as a systematic error.

It has been recently shown by~\citet{DeMattia+2019} that drawing the redshifts of the random catalogues from the data catalogue introduces a radial integral constraint (RIC). We measure the impact of the RIC on the EZmocks by producing a large random catalogue that samples the random catalogues of all the mocks. The observed shift, given in line $\Delta_4$ of Table~\ref{tab:sys-fs}, shows that correction that would need to be applied to correct for the RIC is of the order of 0.7\% on $\alpha_\parallel$, 0.4\% on $\alpha_\perp$ and no effect is seen on $f\sigma_8$. The RIC can be accounted for in the power spectrum model and its effect on cosmological parameter is given in line $\Delta_5$ of Table~\ref{tab:sys-fs}. The agreement with the estimate using different random files is at the per-mil level and we choose, for what follows, to account for the RIC in the model and do not quote a systematic errors for this correction.

The observational systematic errors are summarized in Table~\ref{tab:sys-tot-fs} and summing them in quadrature leads to the following total observational systematic errors:
\begin{equation}
\Delta \alpha_\parallel\bare{obs}=0.005 \quad  \Delta \alpha_\perp\bare{obs}=0.003 \quad\Delta f\sigma_8\bare{obs}=0.004
	\label{eq:fs-obs-syst-errors}
\end{equation}
Adding the modelling error in quadrature (see Table~\ref{tab:sys-tot-fs}), the total systematic errors on the cosmological parameters from the Full Shape RSD Fourier space analysis are:
\begin{equation}
\Delta \alpha_\parallel\bare{syst}=0.011 \quad  \Delta \alpha_\perp\bare{syst}=0.007 \quad\Delta f\sigma_8\bare{syst}=0.012
	\label{eq:fs-syst-errors}
\end{equation}
 These systematic errors represent~30\% of the statistical error and the dominant contribution stems from the modelling of the power spectrum.

\begin{table}
	\centering
	\caption{Systematic errors on the estimate of the cosmological parameters from the Full Shape RSD analysis. The total observational systematic error is the quadratic sum of the errors given in the first rows of the table. Combining in quadrature with the modelling errors determined from the mock challenge gives the total systematic error.}
	\label{tab:sys-tot-fs}
	\begin{tabular}{lccc} 
		\hline
		   & $\alpha_\parallel$ & $\alpha_\perp$ &$f\sigma_8$\\
		\hline
		photometry($\Delta_1$)          & $\pm$0.0030 & $\pm$0.0024 & $\pm$0.0028 \\
        $\Delta(f_s)=10\%$($\Delta_2$)  & $\pm$0.0017 & $\pm$0.0011 & $\pm$0.0021 \\
        fibre collisions ($\Delta_3$)   & +0.0040 & -0.0018 & $\pm$0.0020 \\
		\hline
		total observational & 0.0053  & 0.0032 & 0.0040 \\
		\hline
		\hline
        redshift smearing   & 0.0036 & 0.0042 & 0.0081 \\
	    blind challenge     & 0.0091 & 0.0051 & 0.0093 \\
		\hline
		total modelling     & 0.0098  & 0.0066 & 0.0123  \\
		\hline
		\hline
		total systematics    & 0.0111 & 0.0073 & 0.0129 \\
		statistical error    & 0.0378 & 0.0289 & 0.0447 \\
		fraction             &  30\%  & 25\%  & 29\% \\
		\hline
\end{tabular}
\end{table}

\begin{figure*}
	\includegraphics[width=0.8\paperwidth]{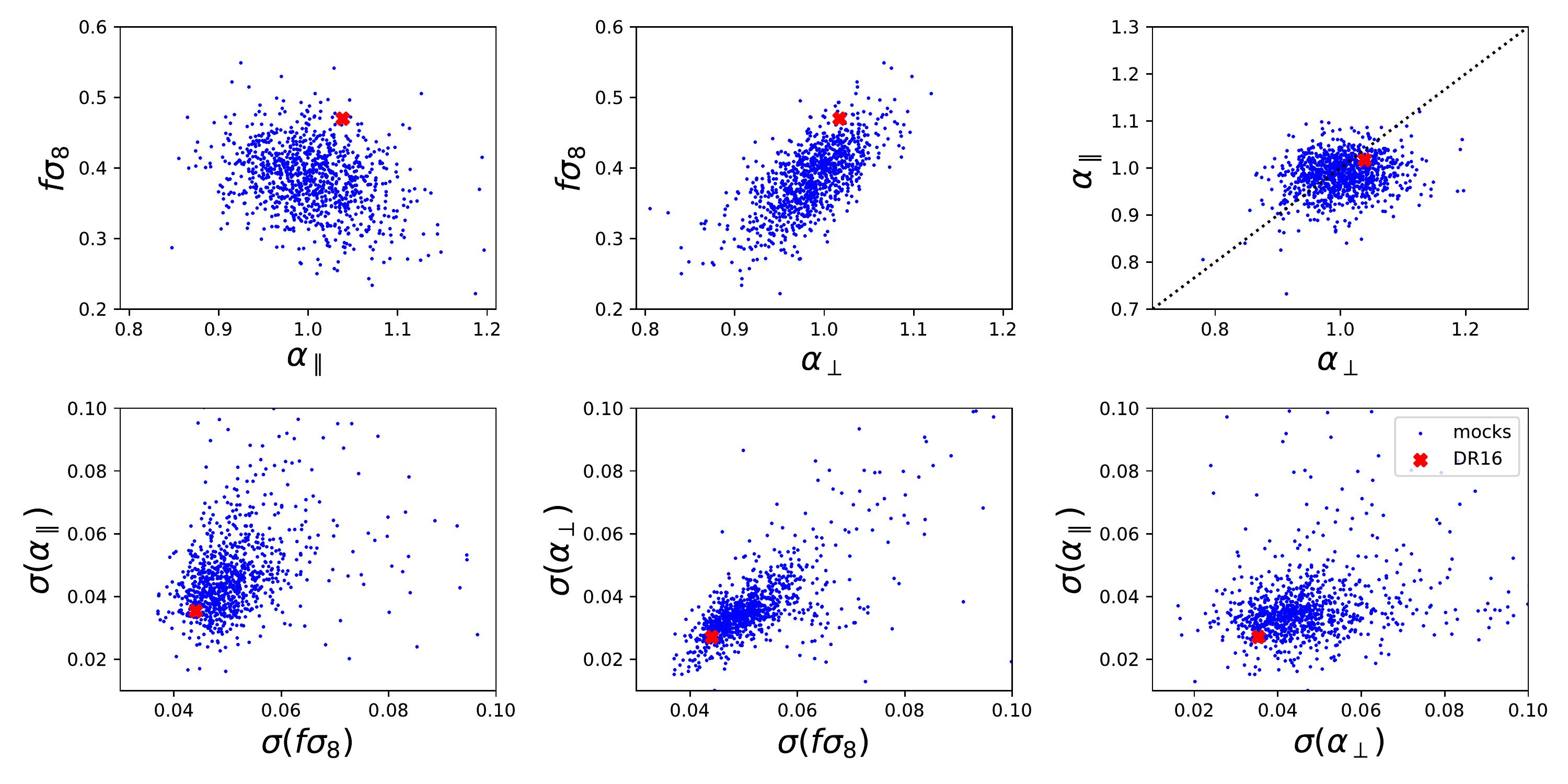}
    \caption{Comparison of the fit parameters and of their errors as measured for the 1000 approximate EZmocks (blue points). The parameters and errors measured for the DR16 sample are represented by a red cross. }
    \label{fig:scatter-FS}
\end{figure*}

In Figure~\ref{fig:scatter-FS}, we show the fit parameters and their errors as measured for the 1000 approximate EZmocks, compared to the DR16 result. Similarly to what was observed for the BAO analysis, the precision of the DR16 sample for the Full-Shape analysis is untypical of the EZmocks and is among the 1\% of mocks with the smallest errors. The interpretation of this is that it comes from the fact that the strength of the BAO is weaker in the EZmocks than in the data. We recall here that the agreement between the EZmocks and the DR16 data for the power spectrum is at the level of a few percent. Therefore the covariance matrix that is used in the fit is correct to this precision. Furthermore, since the amplitude of the BAO is smaller in the EZmocks than in the data, the systematic effects that were estimated using the EZmocks have a larger dispersion and lead to a conservative estimate of the systematic errors.

\subsubsection{Tests on the DR16 sample}
\label{sec:test-fs}

\begin{figure}
	\includegraphics[width=\columnwidth]{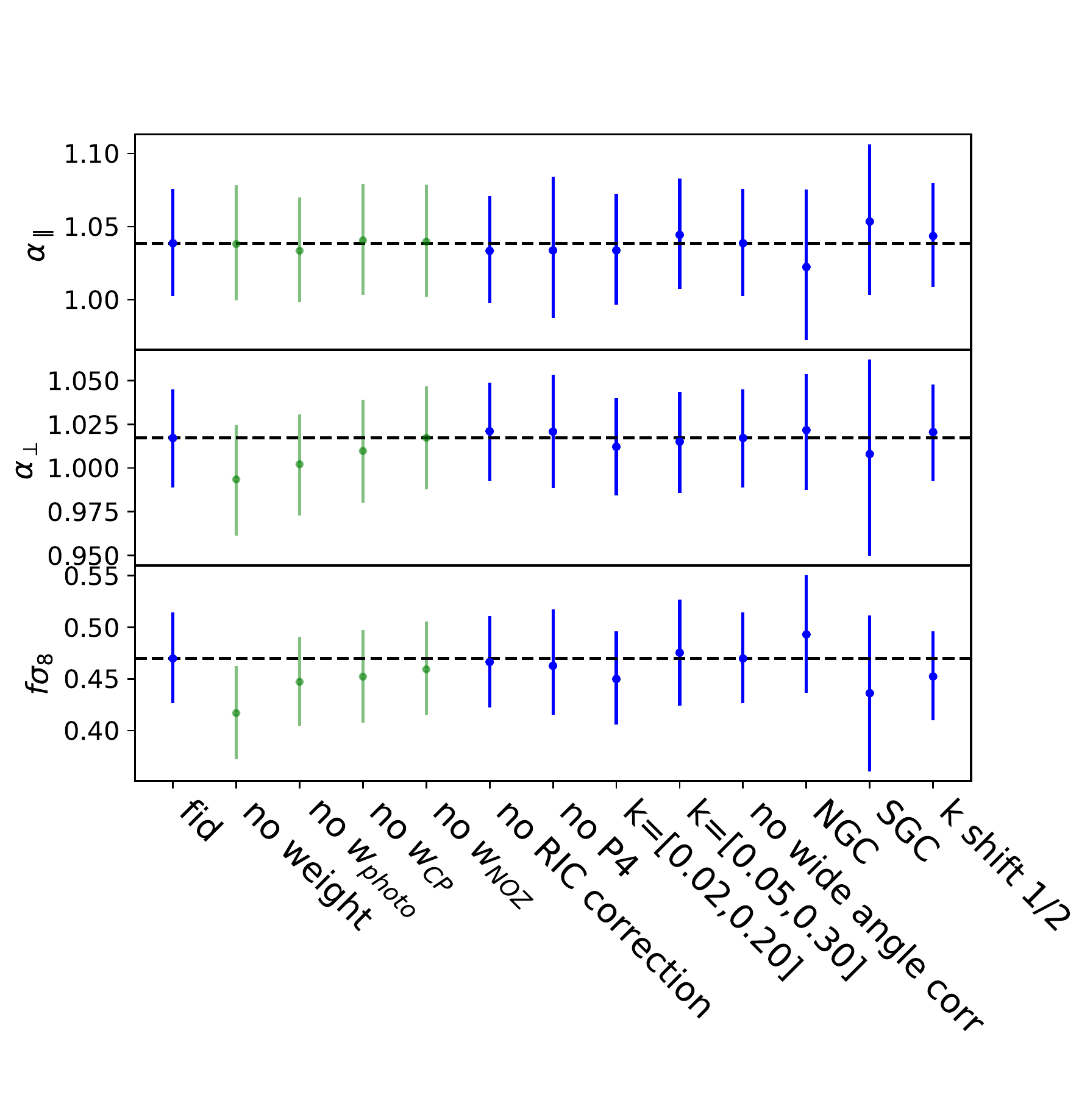}
    \caption{Cosmological parameters measured using the DR16 sample under different choices in the FS analysis. Values of the parameters are taken from Table~\ref{results:test-data}. The subset of green points shows the impact of taking into account different combination of weights to illustrate the size of the correction implied by the weighting scheme but should not be taken as a systematic error.}
    \label{fig:test-DR16}
\end{figure}

We also perform tests to quantify the impact of our choices and for robustness tests to the DR16 data catalogue. Results are summarized in Table~\ref{results:test-data} and displayed in Fig.~\ref{fig:test-DR16}. First, we quantify the effect of each weight that are used to mitigate systematic effects (see previous section). Then, we vary the fitting conditions to evaluate the impact of the analysis choices that were made on the final results. When the change in cosmological parameters is significant, we compare it with the RMS of the mock-by-mock differences distributions and demonstrate that no systematic effect is observed beyond statistics.

\begin{table*}
	\centering
	\caption{Best fit and $\chi^2$ for robustness tests on the data for the Full-Shape RSD analysis. When the difference w.r.t. the reference is significant, we indicate in parenthesis () the RMS of the mock-by-mock differences observed in the EZmocks under the same conditions.}
    \label{results:test-data}
	\begin{tabular}{lccccccc} 
     &$\alpha_\parallel$ & $\alpha_\perp$ &$f\sigma_8$ & $b_{1,{\rm NGC}}\sigma_8$ & $b_{1,{\rm SGC}}\sigma_8$ & $\chi^2(n_{\rm dof}$)& red.$\chi^2$\\
\hline

DR16 bestfit & $1.039 \pm 0.033$ & $1.017 \pm 0.025$ & $0.470 \pm 0.042$ & $0.960 \pm 0.041$ & $0.939 \pm 0.040$ & 116.45 (168-13) & 0.75 \\
no weight  & $1.038 \pm 0.041$ & $0.993 \pm 0.033$ & $0.417 \pm 0.046$ & $0.931 \pm 0.040$ & $0.916 \pm 0.038$ & 95.91 (168-13) & 0.62 \\
no $w_{\rm sys}$  & $1.034 \pm 0.034$ & $1.002 \pm 0.031$ & $0.447 \pm 0.044$ & $0.960 \pm 0.036$ & $0.946 \pm 0.034$ & 115.38 (168-13) & 0.74 \\
no $w_{\rm cp}$  & $1.041 \pm 0.033$ & $1.010 \pm 0.033$ & $0.452 \pm 0.045$ & $0.949 \pm 0.038$ & $0.930 \pm 0.044$ & 104.61 (168-13) & 0.67 \\
no $w_{\rm noz}$  & $1.040 \pm 0.045$ & $1.017 \pm 0.021$ & $0.459 \pm 0.043$ & $0.944 \pm 0.046$ & $0.916 \pm 0.046$ & 106.53 (168-13) & 0.69 \\
no RIC correction  & $1.034 \pm 0.035$ & $1.021 \pm 0.027$ & $0.466 \pm 0.043$ & $0.961 \pm 0.041$ & $0.938 \pm 0.042$ & 117.00 (168-13) & 0.75 \\
no $\ell=4$  & $1.034 \pm 0.056$ & $1.021 \pm 0.038$ & $0.463 \pm 0.055$ & $0.965 \pm 0.035$ & $0.942 \pm 0.043$ & 70.51 (112-13) & 0.71 \\
      & $(0.052)$ & $(0.048)$ & $(0.039)$ & & & & \\
k=[0.02,0.20]  & $1.034 \pm 0.043$ & $1.012 \pm 0.024$ & $0.450 \pm 0.043$ & $0.968 \pm 0.040$ & $0.946 \pm 0.046$ & 82.74 (108-13) & 0.87 \\
& $(0.017)$ & $(0.013)$ & $(0.032)$ & & & & \\
k=[0.05,0.30]  & $1.044 \pm 0.060$ & $1.015 \pm 0.030$ & $0.475 \pm 0.058$ & $0.969 \pm 0.044$ & $0.931 \pm 0.059$ & 104.57 (150-13) & 0.76 \\
     & $(0.018)$ & $(0.015)$ & $(0.020)$ & & & & \\
k shift 1/2  & $1.044 \pm 0.040$ & $1.020 \pm 0.028$ & $0.452 \pm 0.045$ & $0.975 \pm 0.052$ & $0.954 \pm 0.037$ & 125.21 (168-13) & 0.81 \\
     & $(0.019)$ & $(0.017)$ & $(0.019)$ & & & & \\
no wide angle corr  & $1.039 \pm 0.033$ & $1.017 \pm 0.025$ & $0.470 \pm 0.042$ & $0.960 \pm 0.041$ & $0.939 \pm 0.040$ & 116.45 (168-13) & 0.75 \\
NGC  & $1.022 \pm 0.047$ & $1.022 \pm 0.037$ & $0.493 \pm 0.062$ & $0.942 \pm 0.054$ & $-$ & 63.91 (84-8) & 0.84 \\
SGC & $1.054 \pm 0.040$ & $1.008 \pm 0.041$ & $0.436 \pm 0.064$ &  $-$ & $0.952 \pm 0.044$ & 52.05 (84-8) & 0.68 \\
\hline
\end{tabular}
\end{table*}

In Table~\ref{results:test-data} (see also a graphical representation of these results in Fig.~\ref{fig:test-DR16}), we present a series of tests that were performed on the data to evaluate the impact of the choices that were made in the analysis, and the robustness of our measurement. First, we see that applying the complete weighting scheme changes the result of the fit by $\mathcal{O}(1 \sigma)$ for $\alpha_\perp$ and $f\sigma_8$ and has a very small effect on $\alpha_\parallel$. Changing the weighting scheme by removing one of the weights shows that photometric weights ($w_{\rm sys}$) and fibre collisions ($w_{\rm cp}$) have the largest effect on the final results. 

The analysis is also performed by not including the hexadecapole contribution into the fit. As expected, the errors on the parameters increase and the variations of the central values are at most 1/4 of the statistical error. Comparing this to the the RMS of the mock-by-mock differences (in parenthesis in Table~\ref{results:test-data}) shows that the observed shift are within statistics.

Then, we study the stability of the results while changing the boundaries of the k-range or shifting the centre of the bins in $k$ by one half of the bin size. We find that the effect on the dilation scales of the order of $\pm 0.005$ and that there is a substantial effect on $f\sigma_8$ that reaches $\pm 0.019$. Again, the observed shifts are at the level of 1 standard deviation (or less) of the results obtained from the mock-by-mock differences and no additional systematic error is quoted for these effects.

Additional tests were performed with modification made to the modelling. The fit was performed using the modelling of the wide-angle correction as proposed by~\citet{Beutler_2019}, and no difference was observed at a level of precision of 1 per mil. Furthermore, the fit was run using a Gaussian prior of mean $0$ and standard deviation $0.01$ on the quasar-count stochastic term, $A_g$, described in~\ref{sec:model}. The change of cosmological parameters induced is at the level of one tenth of the statistical precision.

The analysis was also performed for the Northern and Southern galactic caps separately and the differences are within 1 standard deviation for each of the cosmological parameters.

\subsubsection{Results from Full Shape RSD analysis}

\begin{figure}
	\includegraphics[width=\columnwidth]{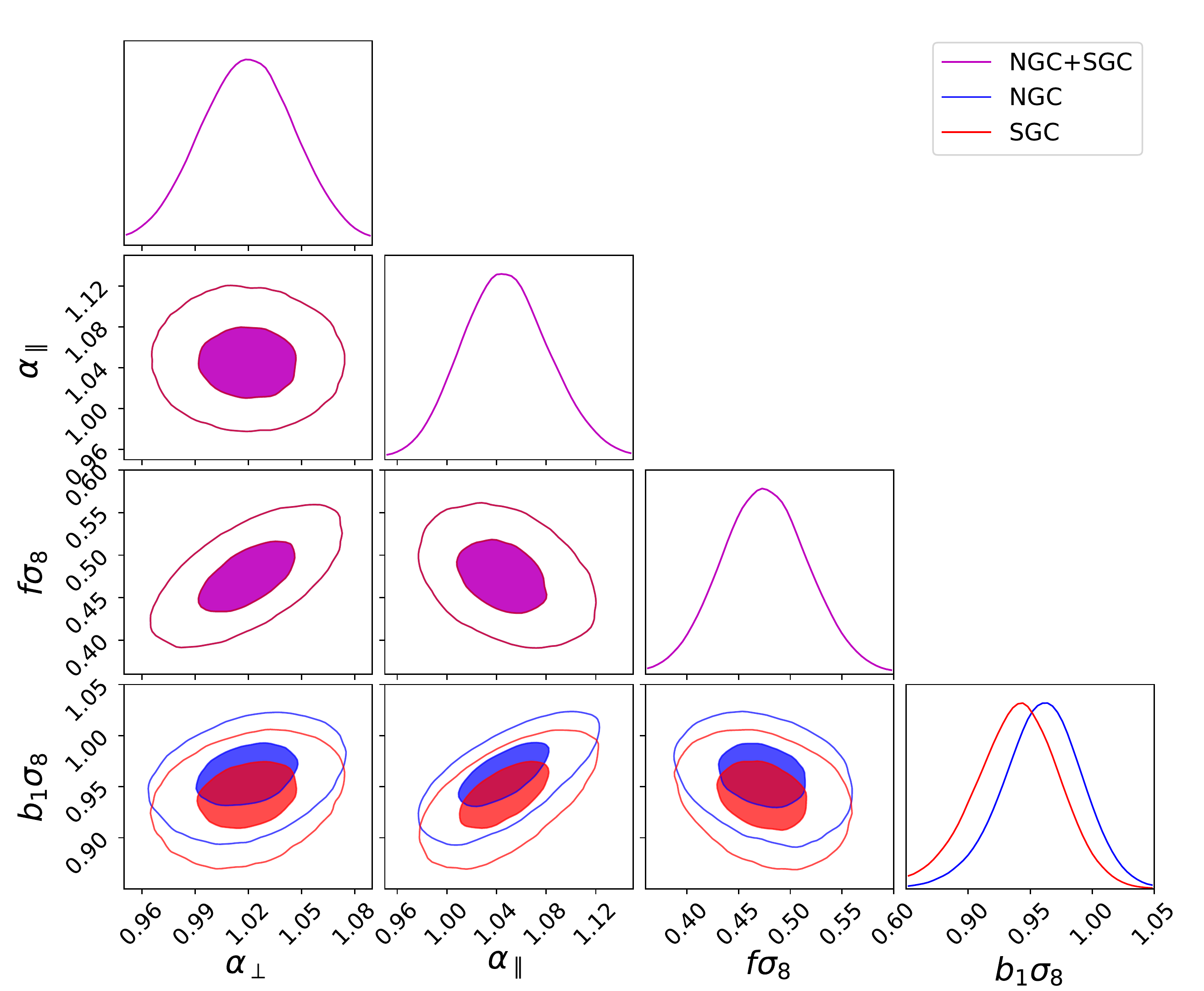}
    \caption{Posterior contour FS for the combined NGC+SGC where the $b_1\sigma_8$ is dependant of the galactic cap}
    \label{fig:scatter-bao-rsd}
\end{figure}

The best-fit Full Shape power spectrum model, for both caps compared to the data, is shown in Figure~\ref{fig:ps-ngc-sgc}. We transform the dilation scale $\alpha_\parallel$ and $\alpha_\perp$ to, respectively, the Hubble distance and the comoving angular diameter distance.
The measurements of the Hubble distance $D_{\rm H}/r_{\rm drag}$, the comoving angular diameter distance $D_{\rm M}/r_{\rm drag}$ and growth rate of structure $f\sigma_8$ from this analysis, are given in Table~\ref{tab:results-ps}. 
The 68\% and 95\% confidence level posterior contours of the cosmological parameters, obtained with a Monte Carlo Markov Chain method are presented in Fig.~\ref{fig:scatter-bao-rsd}. The contours for all possible pairs of parameters including bias and nuisance parameters are given in Appendix~\ref{sec:contours}. The linear bias is allowed to take different values for the two galactic caps, and both values obtained are in agreement.

\begin{table}
    \centering
	\caption{Summary of the results on the Hubble distance $D_{\rm H}/r_{\rm drag}$, the transverse comoving diameter distance $D_{\rm M}/r_{\rm drag}$ and of the linear growth rate of structure $f\sigma_8$. The quoted error is the quadratic sum of the statistical (standard deviation of chains) and systematic errors. The (OR) line shows the results with a fiducial cosmology being the cosmology used for the OuterRim box, see Eq.~\ref{eq:cosmo_OR}}
	\label{tab:results-ps}
	\begin{tabular}{llll} 
     &$D_{\rm H}/r_{\rm drag}$ & $D_{\rm M}/r_{\rm drag}$ &$f\sigma_8$ \\
\hline
DR16         & $13.52\pm{0.51}$ & $30.68\pm{0.90}$ & $0.476\pm{0.047}$\\
DR16 (OR)    & $13.81\pm{0.52}$ & $30.99\pm{0.92}$ & $0.477\pm{0.045}$\\
DR14         & $12.8\pm{0.9}$ & $31.0\pm{1.8}$ & $0.425\pm{0.077}$\\
error ratio & $1.8$ & $2$ & $1.7$\\
\hline
\end{tabular}
\end{table}

To test that the results do not depend on the assumed fiducial cosmology, the complete analysis is also done using the OuterRim cosmology as the fiducial cosmology. The results are in agreement, and the observed differences are comparable with what is calculated from the approximate mocks. The effect of the fiducial cosmology is already included in the systematic errors arising from the modelling as studied in the mock-challenge~\citep{Smith+20} and we do not quote an additional systematic error from the fiducial cosmology at this stage.

The measurement of the linear growth rate of structures is given in term of $f\sigma_8$ and for the linear power spectrum used in the present analysis we have $\sigma_8=0.401$. It is proposed in~\citet{Gilmarin+20}, to use the isotropic dilation scale $\alpha_{\rm iso}=(\alpha_{\parallel}^2\alpha_{\perp})^{1/3}$ to calculate $\sigma_8$ in the cosmology implied by the data. This would decrease our measurement of $f\sigma_8$ by 2.1\% that is close to the systematic error quoted for this parameter. But, changes of cosmologies that could lead to such an effect have already been included in the determination of the systematic errors arising from the modelling. Correcting $\sigma_8$ should in principle also be applied to the mock challenge and would reduce the systematic error, but we leave this for further work. 
In another approach,~\citet{Sanchez2020} proposes to use $\sigma_{12}$ where fluctuations of the linear power spectrum are calculated in spheres of 12$\Mpc$ instead of 8$\Mpch$. Given the value of $h=0.676$ of the fiducial cosmology, the numerical value of $\sigma_{12}$ is only 0.8\% smaller than $\sigma_8$. For completeness, results using this approach are given in appendix D of~\citet{Hou+20}.

Our results are also compared to those obtained for the Fourier space analysis of the eBOSS quasar sample from an earlier data release~\citep[DR14][]{Hector}. The interpretation performed in this previous analysis used a different definition of the effective redshift yielding $z_\mathrm{eff}=1.52$. We recalculate the cosmological parameters $D_{\rm H}(z_{\rm eff})/r_{\rm drag}$ and  $D_{\rm M}(z_{\rm eff})/r_{\rm drag}$ for the DR14 results using our estimate of the effective redshift ($z_\mathrm{eff}=1.480$) and we assume that the 2 samples have the same redshift distribution. The results, given in Table~\ref{tab:results-ps}, show that the results of the two analyses are statistically compatible at 1-sigma level and that the errors are improved by a factor of ~2 for each cosmological parameter using the new data.

\begin{figure}
	\includegraphics[width=\columnwidth]{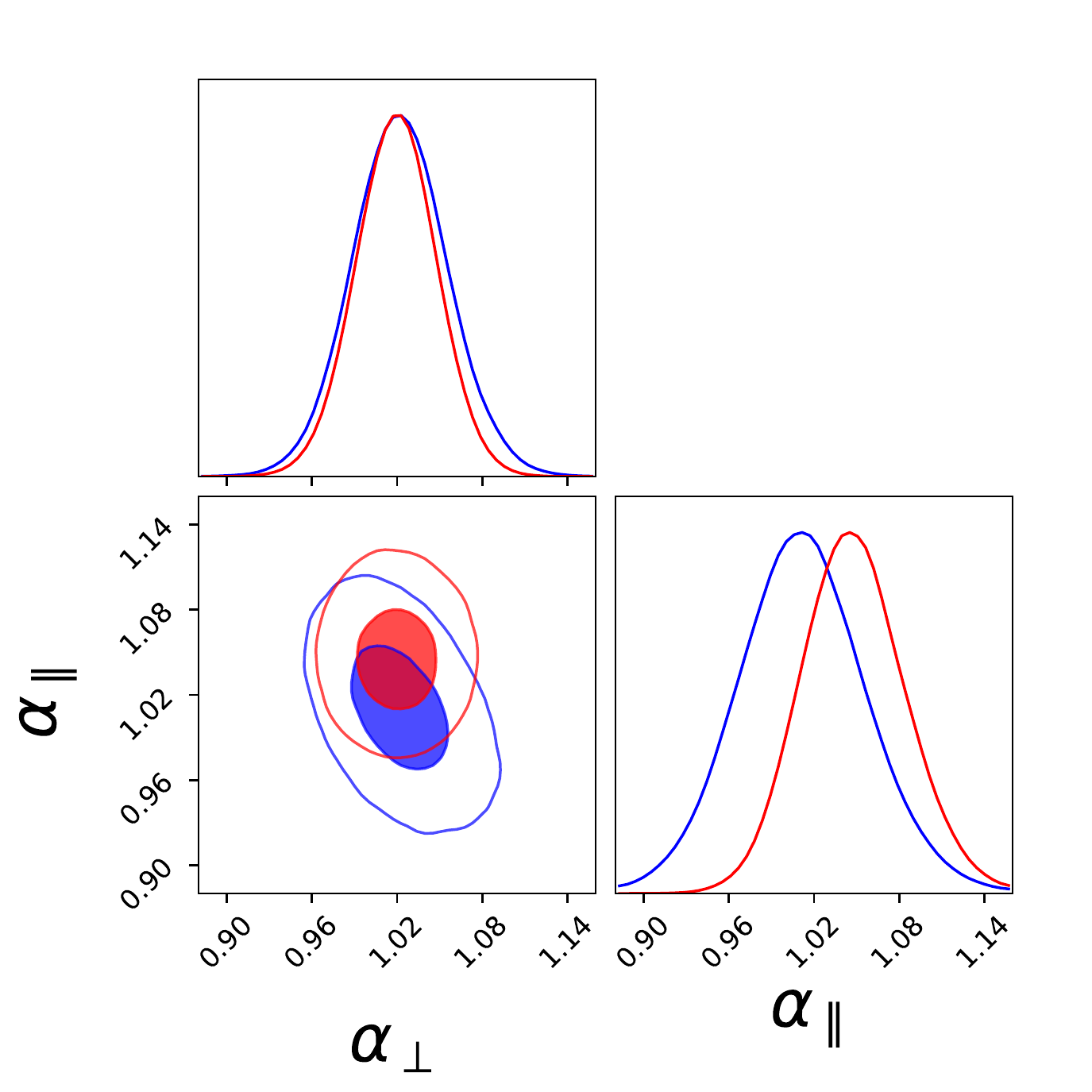}
    \caption{Posterior contours for the BAO (blue) and FS (red) analysis determined with the MCMC chains.}
    \label{fig:scatter-bao-fs}
\end{figure}

The 2D contours of the posterior for $\alpha_\parallel$ and $\alpha_\perp$ from the Full Shape RSD analysis are also compared to the contours obtained for the BAO only analysis (Fig.~\ref{fig:scatter-bao-fs}). The agreement for $\alpha_\perp$ (resp. $\alpha_\parallel$) is within 1/10 (resp. 1/2) of the statistical error. 

\subsubsection{Consensus}

\begin{figure*}
    \centering
    \includegraphics[width=0.6\columnwidth]{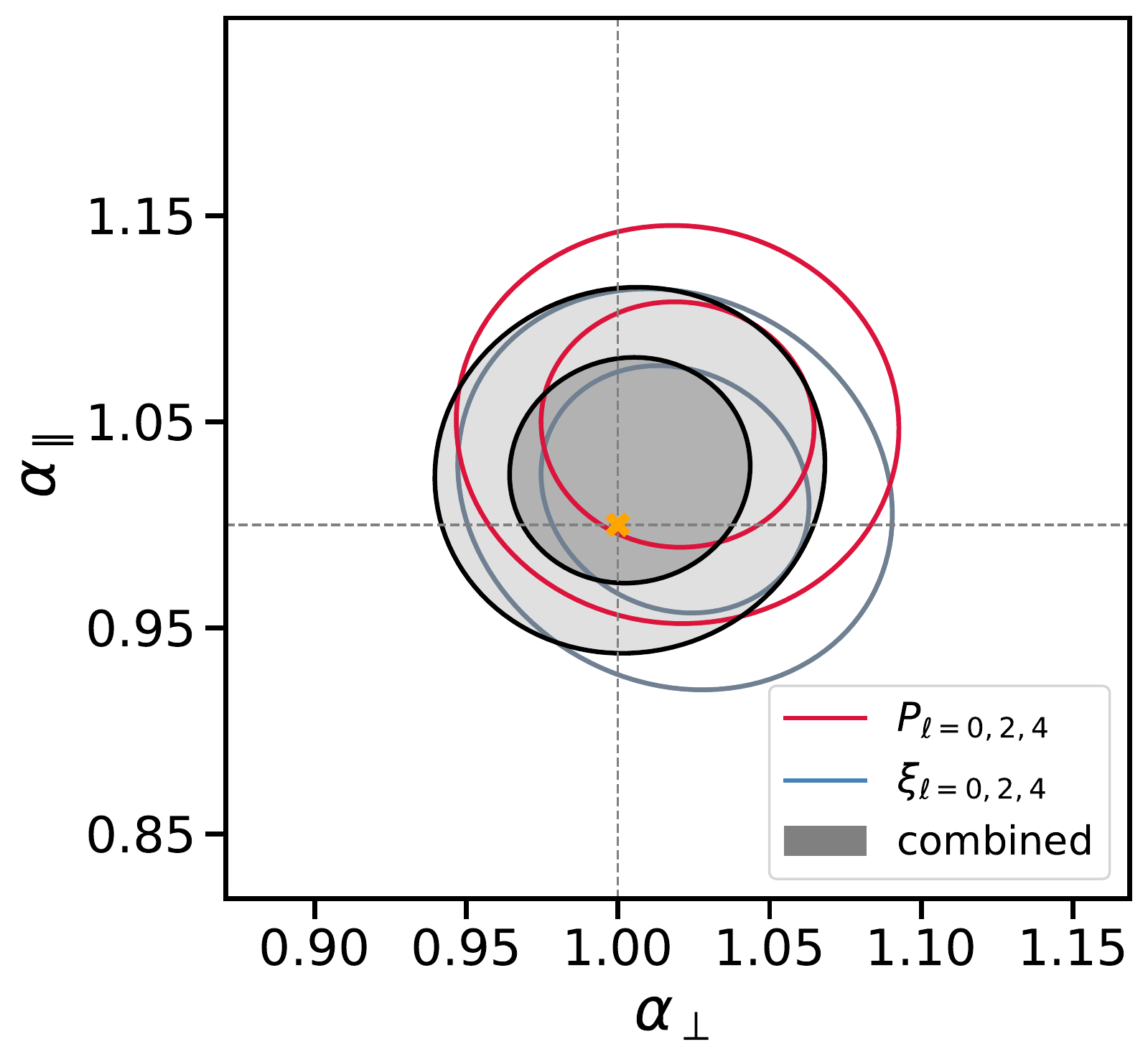}
    \includegraphics[width=0.6\columnwidth]{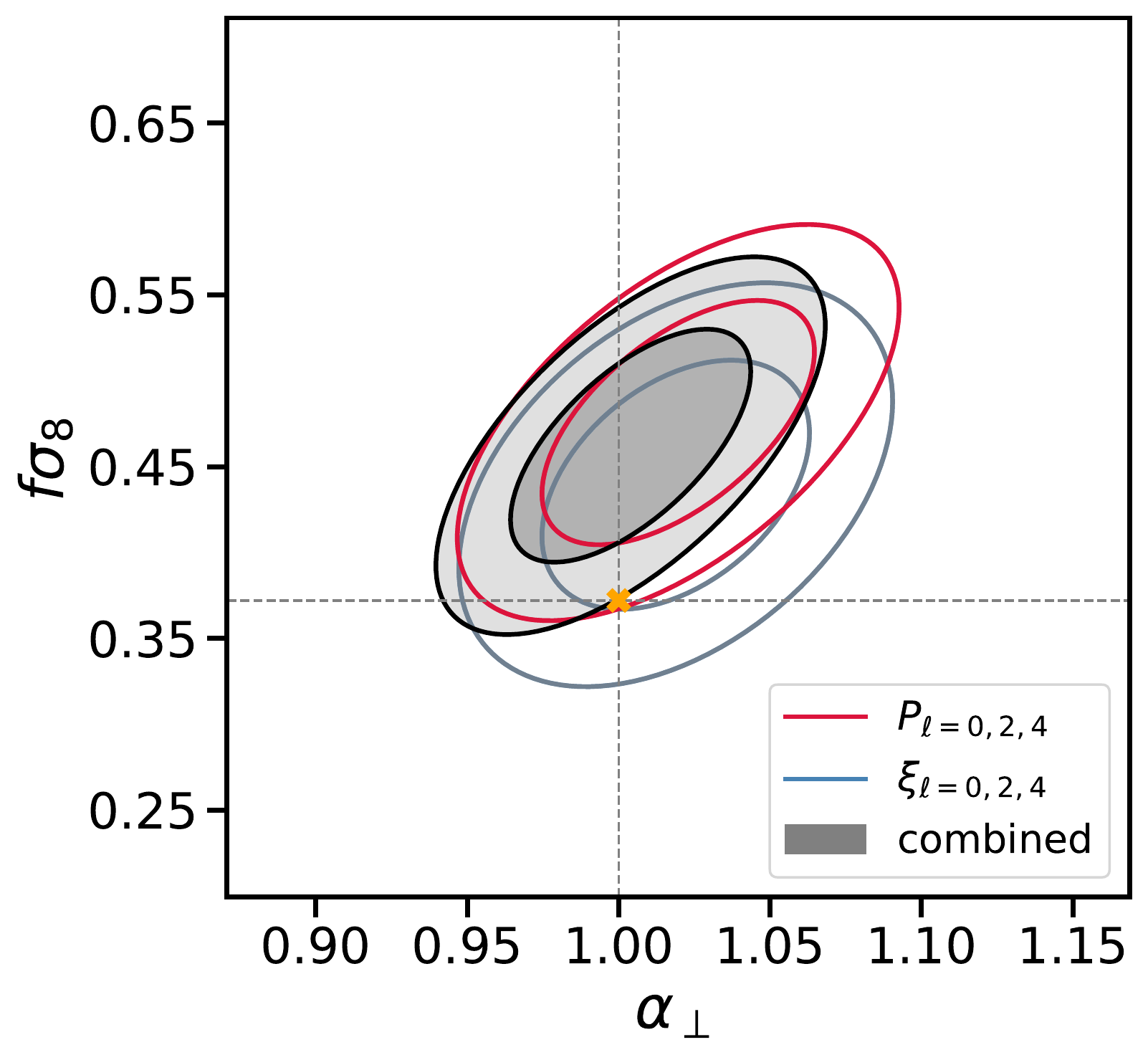}
    \includegraphics[width=0.6\columnwidth]{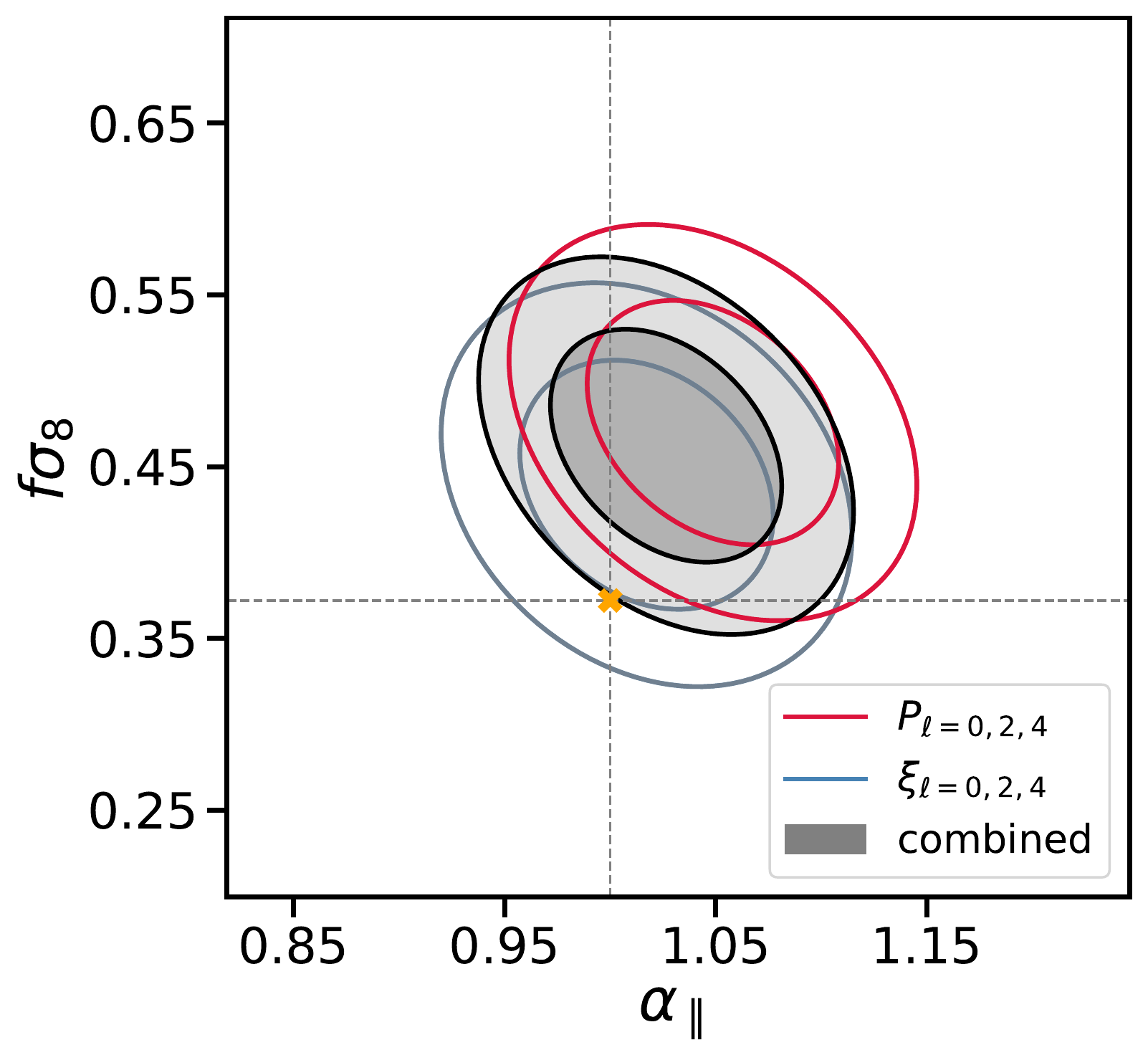}
    \caption{Posterior for $\alpha_{\perp}$, $\alpha_{\parallel}$ and $f\sigma_8$ configuration space, Fourier space and the combined results using the method described in ~\citep{Sanchez+2017}. The filled contours are derived from MCMC chains for configuration space (blue), and k-space(red). The black solid ellipses are the combined constraints at 68, 95 confidence limit. The orange crosses denote the values that are inferred from the combined  Planck2018 and BAO results~\citet{Planck18}.}
    \label{fig:gaussian_posterior_combined_xipk}
\end{figure*}

We perform a consensus analysis of our results with the results obtained in configuration space by~\cite{Hou+20}. The method is based on the work of~\citet{Sanchez+2017} and is described in Section~7.3 of~\cite{Hou+20}. In this method a full $6\times6$ covariance matrix is built from the $3\times3$ covariance matrices of the 2-point correlation function and of the power spectrum measurements, and the cross-terms are determined using the 1000 approximate mocks. The observational systematic errors are added in quadrature to the covariance and we consider that they are independent. The modelling systematic error is determined from the mock challenge, where the consensus technique was applied to each mock realisation and is found to be smaller than either the configuration or Fourier space systematic errors. The results are summarized in Table~\ref{tab:consensus-fullshape} and the posterior contours derived from the MCMC analysis for $\alpha_{\perp}$, $\alpha_{\parallel}$ and $f\sigma_8$ are represented in~Fig.~\ref{fig:gaussian_posterior_combined_xipk}. The measurements are in agreement, and the gain in precision from the consensus is modest. The measurements of $\alpha_{\parallel}$ and $\alpha_{\perp}$ are found to be within 1-$\sigma$ of a flat $\Lambda$CDM model using the cosmological parameters of the combined CMB+BAO measurement of~\citet{Planck18}. Our result of $f\sigma_8$ is 1.9-$\sigma$ above the Planck derived value.

{}

\begin{table*}
\caption{Final results of the Full-Shape analyses in Fourier and configuration spaces and their consensus.}
\begin{center}
\begin{tabular}{|c|c|c|c}
 & $D_{\rm H}(z_{\rm eff})/r_{\rm drag}$ & $D_{\rm M}(z_{\rm eff})/r_{\rm drag}$ & $f\sigma_8$\\
\hline
\hline
Fourier Space & $13.52 \pm 0.51$ & $30.68 \pm 0.90$ & $0.476 \pm 0.047$\\
Configuration Space & $13.11 \pm 0.52$ & $30.66 \pm 0.88$ & $0.439 \pm 0.048$\\
\hline
{\bf Full-Shape Consensus} & \boldmath$13.23 \pm 0.47$ & \boldmath$30.21 \pm 0.79$ & \boldmath$0.462 \pm 0.045$\\
\hline
\end{tabular}
\end{center}
\label{tab:consensus-fullshape}
\end{table*}

\section{Conclusions}
\label{sec:conclusion}

We perform the analysis of the clustering of the complete eBOSS quasar sample. We did two separate analyses. The BAO-only analysis measures the ratio between the angular diameter distance and the sound horizon at the baryon drag epoch, and the ratio between the Hubble distance and the sound horizon at the baryon drag epoch. The Full Shape RSD analysis provides in addition a determination of the linear growth rate of structure times the amplitude of matter density fluctuations. We use a dedicated mock challenge~\citep{Smith+20} to estimate the systematic errors due to the modelling of the power spectrum, and due to the dependence on the assumed fiducial cosmology. The errors due to the observational systematics are determined from approximate mocks where the observational effects have been modelled. For both methods, the dominant source of systematic error resides in the modelling of the power spectrum. The modelling of fibre collisions also has a large impact on the cosmological parameters, especially on the growth rate measurement. The overall systematic errors are at the level of 30\% of the statistical errors. Therefore, improving the models is key for the next generation quasar surveys with increased statistics. A consensus analysis of our measurement in Fourier space and the measurement in configuration space from~\citet{Hou+20} gives the following constraints for the BAO-only analysis:
\begin{equation}
  \begin{split}
   D_{\rm H}/r_{\rm drag} =\frac{c}{Hr_{\rm drag}} &=13.26 \pm 0.55 \\
   D_{\rm M}/r_{\rm drag} &=30.69 \pm 0.80 \\
   \end{split}
   \label{eq:final-bao}
\end{equation}
and for the Full-Shape RSD analysis it yields :
\begin{equation}
  \begin{split}
 D_{\rm H}/r_{\rm drag}    &=13.23 \pm 0.47 \\
 D_{\rm M}/r_{\rm drag} &=30.21 \pm 0.79  \\
 f\sigma_8           &=0.462 \pm 0.045
   \end{split}
   \label{eq:final-fs}
\end{equation}
These measurements are proven very robust by all tests performed. Our measurements of cosmological distances are in agreement with a flat $\Lambda$CDM model using~\citet{Planck18} and our measurement of the linear growth of structures, $f\sigma_8$ is 1.9-$\sigma$ above the Planck derived value. The cosmological interpretation of the DR16 eBOSS quasar sample measurement along with the measurements obtained for the other eBOSS tracers, and the consistency with external data sets are discussed in~\citet{Eva}.

\section*{Acknowledgements}
R. Neveux acknowledges support from grant ANR-16-CE31-0021, eBOSS and from ANR-17-CE31-0024-01, NILAC. 
Funding for SDSS-III and SDSS-IV has been provided by the Alfred P. Sloan Foundation and Participating Institutions. Additional funding for SDSS-III comes from the National Science Foundation and the U.S. Department of Energy Office of Science. Further information about both projects is available at www.sdss.org. SDSS is managed by the Astrophysical Research Consortium for the Participating Institutions in both collaborations. In SDSS-III these include the University of Arizona, the Brazilian Participation Group, Brookhaven National Laboratory, Carnegie Mellon University, University of Florida, the French Participation Group, the German Participation Group, Harvard University, the Instituto de Astrofisica de Canarias, the Michigan State / Notre Dame / JINA Participation Group, Johns Hopkins University, Lawrence Berkeley National Laboratory, Max Planck Institute for Astro- physics, Max Planck Institute for Extraterrestrial Physics, New Mexico State University, New York University, Ohio State University, Pennsylvania State University, University of Portsmouth, Princeton University, the Spanish Participation Group, University of Tokyo, University of Utah, Vanderbilt University, University of Virginia, University of Washington, and Yale University. The Participating Institutions in SDSS-IV are Carnegie Mellon University, Colorado University, Boulder, Harvard-Smithsonian Center for Astrophysics Participation Group, Johns Hopkins University, Kavli Institute for the Physics and Mathematics of the Universe Max-Planck-Institut fuer Astrophysik (MPA Garching), Max-Planck-Institut fuer Extraterrestrische Physik (MPE), Max-Planck-Institut fuer Astronomie (MPIA Heidelberg), National Astronomical Observatories of China, New Mexico State University, New York University, The Ohio State University, Penn State University, Shanghai Astronomical Observatory, United Kingdom Participation Group, University of Portsmouth, University of Utah, University of Wisconsin, and Yale University. This research used resources of the Argonne Leadership Computing Facility, which is a DOE Office of Science User Facility supported under contract DE- AC02-06CH11357. This work made use of the facilities and staff of the UK Sciama High Performance Computing cluster supported by the ICG, SEPNet and the University of Portsmouth. This research used resources of the National Energy Research Scientific Computing Center, a DOE Office of Science User Facility supported by the Office of Science of the U.S. Department of Energy under Contract No. DE-AC02-05CH11231.

\section*{Data availability}
The power spectrum, covariance matrices, and resulting likelihoods for cosmological parameters are (will be made) available (after acceptance) via the SDSS Science Archive Server (https://sas.sdss.org/), with the exact address tbd.




\bibliographystyle{mnras}
\bibliography{biblio} 




\appendix

\section{Posterior distributions derived from the MCMC analysis } \label{sec:contours}

\begin{figure*}
	\includegraphics[width=0.8\paperwidth]{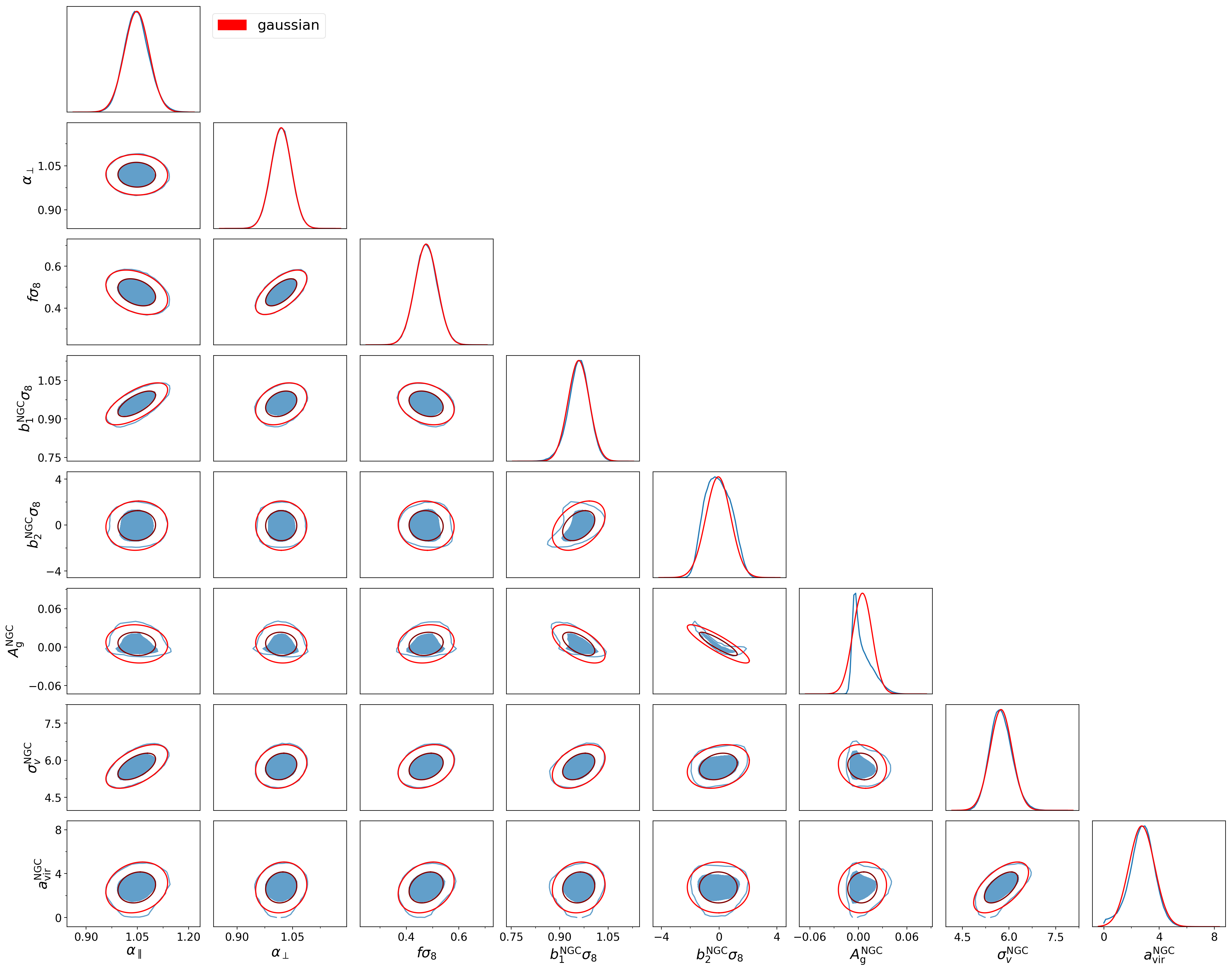}
    \caption{68\% and 95\% confidence level contours (blue) and gaussian approximation (red) for the cosmological parameters and the for the NGC nuisance parameters described in Table~\ref{tab:fs-prior}.}
    \label{fig:results-corner-ngc}
\end{figure*}

\begin{figure*}
	\includegraphics[width=0.8\paperwidth]{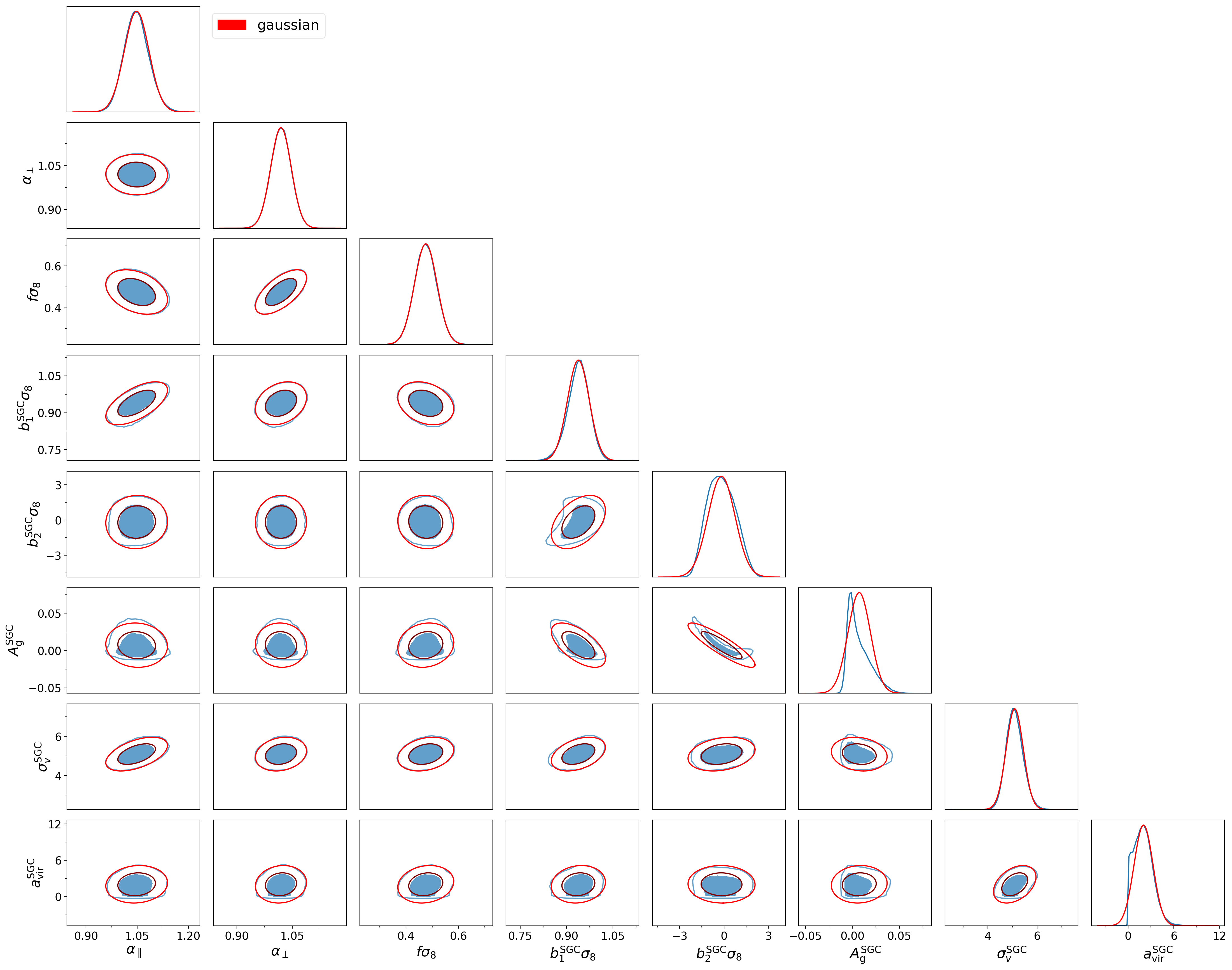}
    \caption{68\% and 95\% confidence level contours (blue) and gaussian approximation (red) for the cosmological parameters and the for the SGC nuisance parameters described in Table~\ref{tab:fs-prior}.}
    \label{fig:results-corner-sgc}
\end{figure*}

In this appendix, we provide the full 68\% and 95\% confidence level contours and Gaussian approximation for all parameters used in the Full-Shape RSD analysis. In Fig.~\ref{fig:results-corner-ngc} (resp. Fig.~\ref{fig:results-corner-sgc}), we show the contours for the cosmological parameters and for the nuisance parameters of Northern (resp. Southern) galactic caps. As stated in the main part of this paper, the contours involving cosmological parameters are found to be Gaussian. For the nuisance parameters, we observe that all contours involving the quasar count stochastic term $A_g$ presented in Section~\ref{sec:FS} are not Gaussian, and that there is strong degeneracy between $A_g$ and the second order bias $b_2$ for both the NGC and the SGC. 

\begin{figure*}
	\includegraphics[width=0.8\paperwidth]{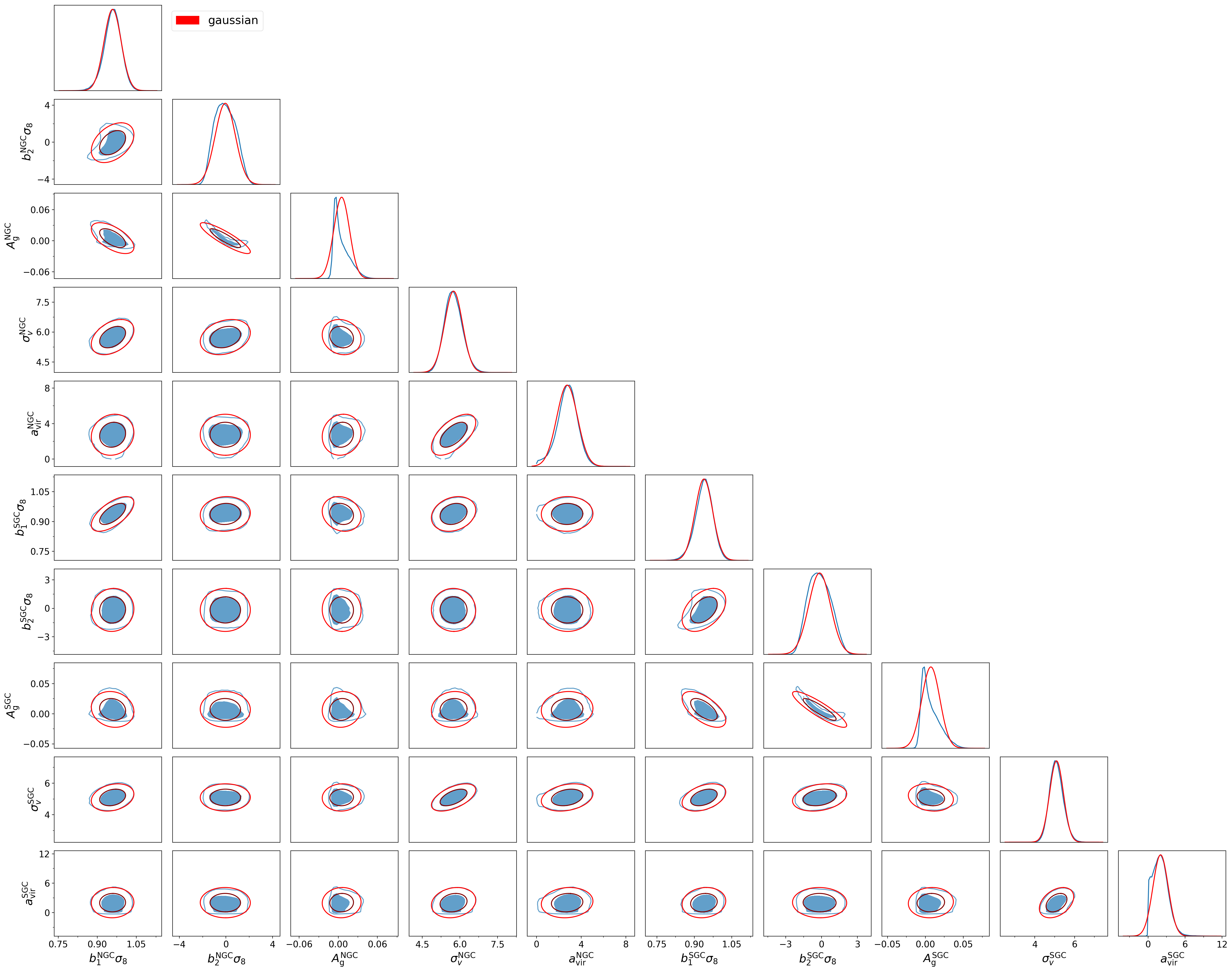}
    \caption{68\% and 95\% confidence level contours (blue) and gaussian approximation (red) for the NGC and SGC nuisance parameters described in Table~\ref{tab:fs-prior}.}
    \label{fig:results-corner-nuisance}
\end{figure*}


\bsp	
\label{lastpage}
\end{document}